\documentclass[preprint,10pt,1p,number]{elsarticle}

\usepackage{amsmath}
\usepackage{amssymb}
\usepackage{hyperref}
\usepackage{subcaption}
\usepackage{dcolumn}
\usepackage{microtype}
\usepackage{url}
\usepackage{lipsum}
\usepackage{physics}
\usepackage{xcolor}
\usepackage{graphicx}

\usepackage{natbib}

\graphicspath{{figures/}}

\newcommand{\pkg}[1]{\texttt{{#1}}}
\newcommand{\cs}{\langle \sigma v \rangle}
\newcommand{\mx}{m_{\chi}}
\renewcommand{\tr}{\tilde{r}}
\newcommand{\tE}{\tilde{E}}

\newcommand{\bb}{b\overline{b}}

\newcommand{\pdm}{\pkg{DarkMatters}}
\newcommand{\prx}{\pkg{RX-DMFIT}}

\journal{Journal of Computational Physics}

\begin{document}

\begin{frontmatter}
  
  \title{DarkMatters: A powerful tool for WIMPy analysis}
  
  \author[a,b]{Michael Sarkis}
  \ead{msarkis@protonmail.com}
  \author[a]{Geoff Beck}
  \ead{geoffrey.beck@wits.ac.za}
  \affiliation[a]{organization={Centre for Astrophysics and School of Physics, University of the Witwatersrand},
            addressline={1 Jan Smuts Ave},
            city={Johannesburg},
            postcode={2101},
            country={South Africa}}
  \affiliation[b]{organization={Department of Physics, Stellenbosch University},
            addressline={111 Merriman Ave},
            city={Stellenbosch},
            postcode={7602},
            country={South Africa}}
  
  \begin{abstract}
    We introduce a new software package, \pdm{}, which has been designed to facilitate the calculation of all aspects of indirect dark matter detection of WIMPs in astrophysical settings. Two primary features of this code are the improvement in performance compared to existing tools, and higher levels of accuracy when determining radio synchrotron emission associated with WIMP annihilations, both of which are enable by the employment of a set of modern and novel numerical techniques. The code also includes functionality for a multi-wavelength set of output products including gamma-ray, radio and neutrino fluxes which can be saved in common formats used by the astronomical community, such as the FITS data file format. The calculations may be tailored to work with a wide range of astrophysical target structures, from dwarf galaxies to galaxy clusters, and the configuration of the underlying calculations is managed by a set of key-value dictionary entries that are easy to understand and use. The code base is publicly accessible through an online repository with a permissive MIT source code licence. 
  \end{abstract}
  

\end{frontmatter}

\section{Introduction}\label{sec:introduction}

The nature of the Dark Matter (DM) in our universe remains a significant unsolved problem in modern physics. The accumulation of evidence over the past several decades, including observations of anisotropies in the Cosmic Microwave Background (CMB)~\cite{planckcollaborationPlanck2018Results2020}, galaxy cluster and merger dynamics~\cite{cloweDirectEmpiricalProof2006}, galactic rotation curves~\cite{rubinRotationAndromedaNebula1970} and gravitational lensing~\cite{masseyDarkMatterGravitational2010}, as well as implications from the current view of the large-scale structure of the universe~\cite{springelLargescaleStructureUniverse2006}, all point toward a likely particle nature of this currently unidentified substance (see~\cite{bertoneHistoryDarkMatter2018} for a recent review). There is thus a global effort underway to try and detect DM, either directly through terrestrial observatories and particle colliders, or through characteristic signatures of astrophysical DM observed indirectly with telescopes. These methods are often complementary in nature, probing different aspects of candidate particle properties, and have grown in scale such that there are now multiple dedicated experiments that generate huge datasets to search through. Of particular interest in this work, the field of indirect detection (ID) presently contains many studies that span multi-wavelength and multi-messenger disciplines, producing constraints on DM model parameters from radio to gamma-ray, comsic ray and neutrino observations. 

Even with the given observational constraints, there are a multitude of viable candidate particle models that have been proposed to fit the role of DM (see~\cite{bertoneParticleDarkMatter2010}). Weakly Interacting Massive Particles (WIMPs) are a generic class of models that contain some neutral, weakly-interacting species with properties that not only provide a clear candidate for the cosmic abundance of DM, but also appear naturally in extensions to the Standard Model (SM) of particle physics. These particles are often considered to be collisionless and cold, due to constraints determined by the layout of large-scale structure~\cite{springelLargescaleStructureUniverse2006,kuhlenNumericalSimulationsDark2012}, which allow them to fit as a component into the standard $\mathrm{\Lambda CDM}$ model of cosmology. The presence of WIMPs within astrophysical structures can be probed through ID, which is predicated on the expectation that they may self-annihilate and produce a set of SM products, which may in turn be observable by existing telescopes and astronomical observatories. 

Given the non-detection of any (confirmed) WIMPs to date, a set of limits on viable particle parameters that generate more indirect emissions than expected have been found (for reviews of the extensive amount of study in this area, see of~\cite{bertoneParticleDarkMatter2010,cirelliDarkMatterIndirect2013,gaskinsReviewIndirectSearches2016}). Prompt gamma-ray fluxes have typically been the most popular search channel in ID studies, with significant interest in the literature propelled by data from the Fermi-LAT space telescope~\cite{fermi/latcollaborationLargeAreaTelescope2009} and a tantalising excess signal found at the galactic centre~\cite{goodenoughPossibleEvidenceDark2009,hooperDarkMatterAnnihilation2011}, which is still generating debate and further investigation~\cite{dimauroCharacteristicsGalacticCenter2021,cholisReturnTemplatesRevisiting2022}. There has also been a significant effort to search the dwarf spheroidal (dSph) satellite galaxies of the Milky Way (MW), as these targets are highly DM-dominated structures with relatively low baryonic foreground emissions, with gamma-ray observations of dozens of dSphs used in recent studies~\cite{hooperGammaRayEmissionReticulum2015,fermi-latcollaborationSearchingDarkMatter2015,hoofGlobalAnalysisDark2020,armandCombinedDarkMatter2021}. Radio wavelength ID studies have also enjoyed recent interest, driven by the excellent observing capabilities of modern radio interferometer telescopes like the LOFAR~\cite{vanhaarlemLOFARLOwFrequencyARray2013}, ATCA~\cite{wilsonAustraliaTelescopeCompact2011}, JVLA~\cite{laneVeryLargeArray2014}, and most recently the SKA precursor instruments like MeerKAT~\cite{jonasMeerKATRadioTelescope2018} and ASKAP~\cite{johnstonScienceASKAPAustralian2008}. Since the reference work of~\cite{colafrancescoMultifrequencyAnalysisNeutralino2006}, the number of radio wavelength ID studies has grown substantially, and now includes notable works regarding dSphs~\cite{regisLocalGroupDSph2015a, regisDarkMatterReticulum2017, vollmannRadioConstraintsDark2020, gajovicWeaklyInteractingMassive2023}, galaxies like M31~\cite{egorovUpdatedConstraintsWIMP2022} and the Large Magellanic Cloud~\cite{regisEMUViewLarge2021}, and galaxy clusters~\cite{stormSynchrotronEmissionDark2017,kiewConstraintsDarkMatter2017,chanPossibleRadioSignal2020,sarkisRadiofrequencySearchWIMPs2023,lavisRadiofrequencyWIMPSearch2023}. Finally, the use of multi-messenger species in ID studies has also advanced substantially in the past several years, with notable studies including~\cite{aartsenSearchNeutrinosDark2017,albertSearchDarkMatter2020,abeIndirectSearchDark2020,abbasiSearchNeutrinoLines2023} for neutrino searches and~\cite{yuanInterpretationsDAMPEElectron2017,yangDarkMatterAnnihilation2017, cholisRobustExcessCosmicRay2019,fanModelExplainingNeutrino2018,beckExcessExcessesExamined2019,caloreAMS02AntiprotonsDark2022} for cosmic ray searches. 

The studies highlighted above display the abundance of both data availability and interest in the literature, and new generations of observatories are likely to only amplify this abundance. To keep up with this incoming data, it is vital for ID studies to have access to a modelling framework that provides the accuracy needed for high-resolution data and the computational efficiency needed for high data volumes. There are existing open-source tools that are able to compute the multi-wavelength emissions from WIMPs, the most notable of which is the \prx{}~\cite{mcdanielMultiwavelengthAnalysisDark2017} package which was developed to provide the earlier \pkg{DM-FIT}~\cite{jeltemaFittingGammaRaySpectrum2008} package with X-ray and radio flux capabilities, and the recent DM-related patches which extend the functionality of the \pkg{GALPROP}~\cite{strongPropagationCosmicrayNucleons1998} package by the author of~\cite{egorovUpdatedConstraintsWIMP2022}\footnote{Patches can be found in \url{https://github.com/a-e-egorov/GALPROP_DM} until they are upstreamed}. The \pkg{GALPROP}~\cite{strongPropagationCosmicrayNucleons1998} and~\pkg{DRAGON(2)}~\cite{evoliCosmicrayPropagationDRAGON22017} code packages provide solutions to the related problem of cosmic-ray transport in the galaxy, and some of the numerical techniques used in these packages have inspired development of segments of the code presented in this work. However, each of these open-source tools lack some functionality in what we would consider as a complete solution for a highly accurate, efficient and generalised ID DM tool. We have thus developed the \pdm{} package to overcome these limitations and we provide the tool in a permissive and open-source format to the community.

The structure of this paper is as follows. Section~\ref{sec:modelling} contains a description of the theoretical framework for all physical models, including DM halo and particle models, gas density and magnetic field profiles, and resulting multi-wavelength emission. Section~\ref{sec:implementation} includes details of how to interface with the code and general outlines for obtaining and using the software. In Section~\ref{sec:comparisons}, a set of comparison calculations with an existing software package, \prx{}, are described and any differences apparent in the results are analysed. Section~\ref{sec:conclusion} contains a discussion on the use of this package in the wider context of astrophysical DM indirect detection, as well as a set of concluding remarks. A detailed technical description of the numerical method used to solve the electron propagation equation is additionally provided in~\ref{sec:solution_methods}.

\section{Modelling framework}\label{sec:modelling}

\subsection{DM halos}\label{sec:dm-halos}

There are several physical characteristics that play a role in the modelling of DM halos. Although there is evidence to suggest that DM halos are likely to have tri-axial shapes (see~\cite{dubinskiStructureColdDark1991,bettSpinShapeDark2007}), we make the common assumption that the halos are spherically symmetric. This assumption correlates with the spherically symmetric magnetic field and gas density profiles used in the code, and should provide a good approximation for many astrophysical scenarios. The full halo parameters can be divided into two categories. The first are the characteristic values of density and radius: $\rho_s$ and $r_s$ respectively. The second set contains the virial mass, radius, and concentration: $M_\mathrm{vir}$, $R_\mathrm{vir}$, and $c_\mathrm{vir}$. The minimum information required to specify a halo is simply a virial mass/radius. This is combined with numerical fitting functions for the concentration to fully specify the halo parameters. Otherwise, the following pairs are accepted as minimum information: ($r_s$, any), ($R_\mathrm{vir}$,$c_\mathrm{vir}$), and ($M_\mathrm{vir}$,$c_\mathrm{vir}$). We will now proceed to specify the definitions of each variable.

First, we note that the virial radius is determined according to
\begin{equation}
   \frac{1}{\frac{4}{3}\pi R_{\mathrm{vir}}^3} \int^{R_\mathrm{vir}}_0 4 \pi r^2 \rho(r) dr = \Delta_c(z) \rho_c(z) \, ,
\end{equation}
with the critical density being given by
\begin{equation}\label{eqn:rho_crit}
	\rho_c(z) = \frac{3H(z)^2}{8\pi G}\,,
\end{equation}
while the virial contrast $\Delta_c$ is 
\begin{equation}\label{eqn:overdensity}
	\Delta_{c} \approx 18\pi^2 - 82x - 39x^2 \; ,
\end{equation}
where $x = 1 - \Omega_m(z)$. Here $\Omega(z)$ is the usual cosmological density parameter, for matter ($m$) or the cosmological constant ($\Lambda$), and $\Omega_m(z)$ is given by
\begin{equation}\label{eqn:omega}
	\Omega_m(z) = \dfrac{1}{1+\frac{\Omega_{\Lambda}(0)}{\Omega_m(0)}(1+z)^{-3}} \, .
\end{equation}
We can then define the virial mass via
\begin{equation}\label{eqn:virial_mass}
	M_{\mathrm{vir}} = \dfrac{4}{3}\pi\Delta_c\rho_cR^3_{\mathrm{vir}} \,,
\end{equation}
and the virial concentration by
\begin{equation}\label{eqn:virial_concentration}
	c_{\mathrm{vir}} = \dfrac{R_{\mathrm{vir}}}{r_{-2}} \,.
\end{equation} 
Note that $r_{-2}$ is the radius where the logarithmic slope of the density profile $\rho(r)$ is equal to $-2$. This is not always equal to $r_s$ but the code accounts for this discrepancy in all of the predefined density profiles. The virial concentration can also be determined by one of a suite of numerical fitting functions from Prada 2012~\cite{pradaHaloConcentrationsStandard2012}, Muñoz-Cuartas 2011~\cite{munoz2011}, Colafrancesco 2006~\cite{colafrancescoMultifrequencyAnalysisNeutralino2006}, or Bullock 2001~\cite{bullockProfilesDarkHaloes2001}. 

The first halo density profile available in the code is the Navarro-Frenk-White (NFW) profile~\cite{navarroStructureColdDark1996}
\begin{equation}\label{eqn:nfw}
	\rho_{\mathrm{nfw}}(r) = \dfrac{\rho_s}{\left( \dfrac{r}{r_s} \right)\left(1+\dfrac{r}{r_s}\right)^{2}} \,.
\end{equation}
There is also a generalised form of this profile
\begin{equation}\label{eqn:gnfw}
	\rho_{\mathrm{gnfw}}(r) = \dfrac{\rho_s}{\left( \dfrac{r}{r_s} \right)^{\alpha}\left(1+\dfrac{r}{r_s}\right)^{3-\alpha}} \,,
\end{equation}
with free parameter $\alpha$.
We also have the Einasto profile~\cite{einastoConstructionCompositeModel1965}
\begin{equation}\label{eqn:einasto}
	\rho_{\mathrm{ein}}(r) = \rho_s \exp\left\{ -\dfrac{2}{\alpha} \left[ \left( \dfrac{r}{r_s} \right)^{\alpha} -1 \right] \right\} \,,
\end{equation}
again with free parameter $\alpha$.
Then there is Burkert's profile~\cite{burkertStructureDarkMatter1995}
\begin{equation}\label{eqn:burkert}
	\rho_{\mathrm{bur}}(r) = \dfrac{\rho_s}{\left( 1 + \dfrac{r}{r_s} \right) \left(1+\left( \dfrac{r}{r_s} \right)^{2}\right) } \,,
\end{equation}
and the pseudo-isothermal profile
\begin{equation}\label{eqn:p-iso}
	\rho_{\mathrm{iso}}(r) = \dfrac{\rho_s}{1+\left( \dfrac{r}{r_s} \right)^{2}} \,.
\end{equation}

\subsection{Particle spectra}\label{sec:particle-spectra}

The observable radiation from the halos described above is also dependent on the nature of the constituent DM particles. The generic WIMP models considered here are parameterised by two quantities: the WIMP mass ($\mx$), and either cross-section ($\cs$) or decay rate ($\Gamma$) in the case of annihilation or decay, respectively. The energy spectrum of the SM particles produced by each WIMP annihilation or decay is then represented by $dN_i/dE$, where $N_i$ is the produced multiplicity, $i$ refers to the species of the final state stable SM particle, and $E$ is the corresponding particle energy. The final state is reached through various intermediate particle interactions -- known as channels and denoted by $f$ -- which can be represented generally through the reaction pathway of $\chi\chi \rightarrow f \rightarrow i$ ($\chi \rightarrow f \rightarrow i$) in the case of annihilation (decay). A general form of the produced energy spectrum is given by 
\begin{equation}\label{eqn:particle_spectrum}
	\dv{N_i}{E} = \sum_f B_f \,\dv{N_{f\rightarrow i}}{E} \,,
\end{equation}
where the considered spectrum is a combination of all relevant intermediate channels, and the contribution of each channel is determined by the branching ratio $B_f$. As is standard in the indirect-detection literature, we consider results for individual channels (corresponding to a branching ratio of $B_f = 1$ for the channel of interest) that can be specified explicitly for each computation within the code (See Section~\ref{sec:input_files}). The final state species are likewise chosen according to the desired observable to be computed ($i$ would be gamma-ray photons for the case of calculating prompt gamma-ray emissions, for example), and these will be discussed in further detail in Section~\ref{sec:observables}. 

The produced particle spectra, as calculated from any source, can be specified as an input to the code. By default, \verb|DarkMatters| makes use of the latest version of the pre-computed and model-independent numerical tables provided by the Poor Particle Physicist Cookbook for Dark Matter Indirect Detection (PPPC4DMID)~\cite{cirelliPPPCDMID2011}, which are also hosted in an online repository~\footnote{\url{http://www.marcocirelli.net/PPPC4DMID.html}}. The tables (`ingredients') provided by the PPPC4DMID have also been computed with corrections for electroweak radiative effects, following the prescription given in~\cite{ciafaloniWeakCorrectionsAre2011}. These corrections can have an impact on the final energy spectrum, especially when particle energies are larger than the electroweak scale, which is relevant for DM models with a large WIMP mass. The accessible and model-independent nature of the results provided by the PPPC4DMID allows for a convenient utilisation of these particle spectra, which has made this package a popular resource in recent indirect-detection literature.

To then determine the spatial and energy distribution of the particles that are injected into the halo from WIMP annihilation and decay,  the particle spectrum defined above is multiplied by the number density of WIMPs to yield the source function $Q(r,E)$, which can be written as follows:
\begin{equation}\label{eqn:source_general}
Q(r,E) = 
    \begin{cases}
        \displaystyle \dfrac{1}{2}\left(\dfrac{\rho_{\chi}(r)}{\mx}\right)^2 \cs \dv{N_{i}}{E}   \,, &\qquad\text{(annihilation)} \\[1em]
        \displaystyle \left(\dfrac{\rho_{\chi}(r)}{\mx}\right) \Gamma \,\dv{N_{i}}{E}   \,. &\qquad\text{(decay)} 
    \end{cases}
\end{equation}
In these equations, $\rho_{\chi}$ represents the radial density of WIMPs in the halo as described by the profiles given in the previous section. The number density in the case of annihilation assumes Majorana WIMP pairs, and one should include an additional factor of 1/2 for the case of Dirac WIMPs.  

\subsection{Electron propagation}\label{sec:propagation}

Following the annihilation or decay of WIMPs, the indirect detection of the observable products will depend on the final state of the SM particles. In the case of gamma-ray photons or neutrinos, their propagation from the DM halo to the Earth is relatively straightforward, as they do not interact with intermediate environments like magnetic fields (see Sections~\ref{sec:gamma-ray} and~\ref{sec:neutrino}). However, for radio-frequency synchrotron emissions (resulting from $\chi\chi \rightarrow f \rightarrow e^{\pm}$), the evolution of the charged final state electrons is goverened by several physical interactions between the electrons and magnetic fields and thermal gas populations that are ubiquitous in large astrophysical structures. To model these effects, one typically employs a cosmic-ray transport equation, such as the one used in the \verb|GALPROP| package~\cite{porterGALPROPCosmicrayPropagation2022}, and presented in the Appendix of~\cite{strongPropagationCosmicrayNucleons1998}. Although this form encapsulates all possible physical interactions, the two dominant effects in most astrophysical scenarios that pertain to the indirect detection of WIMPs in large DM halos are those of spatial diffusion and energy losses. By neglecting all sub-dominant terms, the full transport equation reduces to the standard diffusion-loss equation, 
\begin{align}\label{eqn:diffusion_loss}
	\pdv{\psi(\mathbf{x},E)}{t} = \nabla\cdot(D(\mathbf{x},E)\nabla\psi(\mathbf{x},E)) + \pdv{}{E}\left(b(\mathbf{x},E)\psi(\mathbf{x},E)\right) + Q(\mathbf{x},E) \, .
\end{align}
In this equation, we have represented the spatial and energy distribution of electrons by $\psi$, as in \pkg{Galprop}. We further simplify this equation by assuming spherical symmetry, so that $\mathbf{x}\rightarrow r$, which should be valid for typical cool-core galaxy clusters and dSphs (for an example of the detailed solution to this equation without this assumption in the M31 galaxy, see~\cite{egorovUpdatedConstraintsWIMP2022,egorovNatureM31Gammaray2023a}). The solution to this partial differential equation is the equilibrium electron distribution, which can be found through one of several mathematical techniques (this is discussed in further detail in~\ref{sec:solution_methods}).

The effects of diffusion and energy loss are determined by the corresponding coefficients in Equation~\eqref{eqn:diffusion_loss}, $D(r,E)$ and $b(r,E)$, respectively. In \pkg{DarkMatters}, we utilise the following general form of the diffusion coefficient,
\begin{equation}\label{eqn:diffusion}
	D(r,E) = D_0\left(\frac{E}{1\,\mathrm{GeV}}\right)^{\delta}\left( \frac{B(r)}{B(0)} \right)^{-\delta} \,,
\end{equation}
where $D_0$ is the diffusion normalisation constant, $\delta$ is the diffusion index and $B(r), B_0$ represent the magnetic field profile and normalisation, respectively. When modelling the effects of spatial diffusion with Equation~\eqref{eqn:diffusion}, it is necessary to choose a description of the turbulence in the magnetic field that gives rise to the diffusion of the charged particles in that field. This choice determines the value of the index $\delta$, and common descriptions include the cases of Kolmogorov ($\delta = 1/3$), Kraichnan ($\delta = 1/2$), and Bohm-like ($\delta = 1$) diffusion. The factor of $(B(r)/B_0)^{-\delta}$ in Equation~\eqref{eqn:diffusion} traces the radial profile of the magnetic field, which results in an equilibrium electron distribution that accurately reflects the geometry of the environment when using numerical solution methods, and a form of this factor has been used before in the works by~\cite{regisDarkMatterReticulum2017,beckGalaxyClustersHigh2023}. 

The energy-loss coefficient, defined as the energy-loss rate $b(r,E) \equiv \dd E/\dd t$, encapsulates all of the physical process that reduce the total energy of the electron distribution. The dominant effects at high energies (relevant for typical WIMP models) are synchrotron emission and Inverse Compton (IC) scattering, though we also include the effects of Coulomb scattering and bremsstrahlung, which can be relevant at lower energies. The full form of the energy-loss coefficient that is used in \pkg{DarkMatters} when solving Equation~\eqref{eqn:diffusion} is given below, 

\begin{align}\label{eqn:energy_loss}
	b(r,E) &= b_{\text{IC}}\left( \frac{U_{ph}}{\mathrm{eV/cm^3}}\right)\left(\dfrac{E}{1\,\mathrm{GeV}}\right)^2 \nonumber \\
	&+ b_{\text{synch}}\left(\dfrac{E}{1\,\mathrm{GeV}}\right)^2\left(\dfrac{B(r)}{1\,\mu\mathrm{G}}\right)^2 \nonumber\\
	&+ b_{\text{Coul}}\left(\dfrac{n_e(r)}{1\,\mathrm{cm^{-3}}}\right)\left(\log\left(\dfrac{\gamma}{n_e(r)/\mathrm{1\,cm^{-3}}}\right) + 73\right) \nonumber\\
	&+b_{\text{brem}} \left(\dfrac{n_e(r)}{1\,\mathrm{cm^{-3}}}\right) \gamma \left(\log\left(\gamma\right)+0.36\right) \,,
\end{align}

where $n_e(r)$ is the thermal gas number density, $U_{ph}$ is the target photon energy density ($U_{ph} = 0.26\, \mathrm{eV/cm^3}$ in the case of CMB photons) and $\gamma$ is the usual Lorentz factor. Each of the effect's loss rate coefficients are taken from the values presented in~\cite{egorovUpdatedConstraintsWIMP2022}, which have recently been corrected from errors in previous works. The coefficients, in units of $\mathrm{GeV\,s^{-1}}$, are thus $ b_{\text{IC}} = 1.0 \times 10^{-16}\;, b_{\text{synch}} = \; 0.025 \times 10^{-16},\; b_{\text{Coul}} = 7.6 \times 10^{-18},\; b_{\text{brem}}  =  7.1 \times 10^{-20}$.  

\begin{figure}[htbp]
  \centering
  \includegraphics[width=0.75\linewidth]{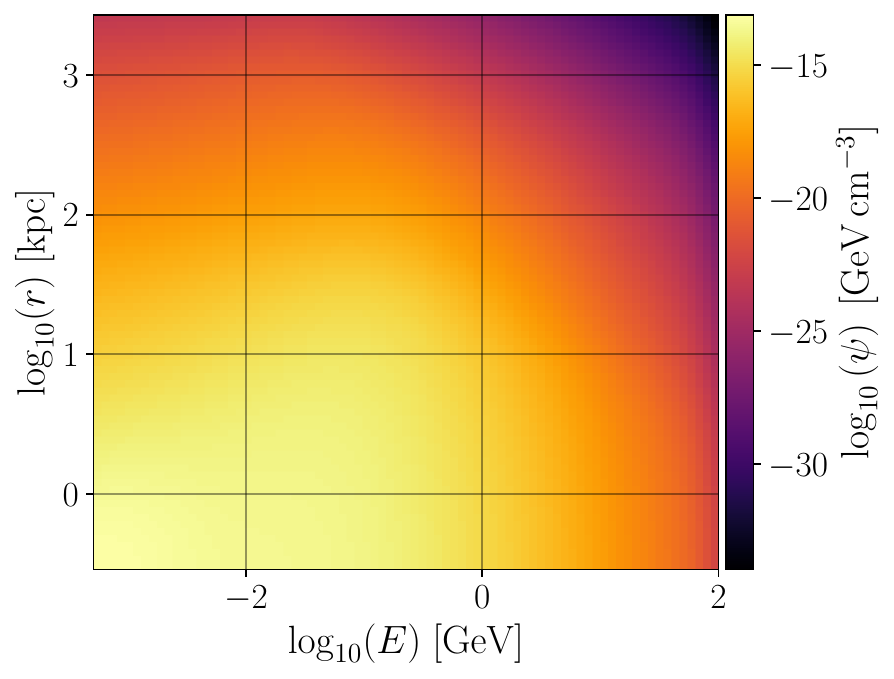}
  \caption{The equilibrium distribution of electrons, computed numerically as the solution to the diffusion-loss equation (Equation~\eqref{eqn:diffusion_loss}). This solution is for a set of sample DM halo and particle parameters, namely for the Coma galaxy cluster with an NFW halo profile and 100 GeV WIMPs annihilating solely through the $\chi\chi\rightarrow \bb\rightarrow e^{\pm}$ channel.}
  \label{fig:coma_electrons}
\end{figure}

With the explicit forms of the functions $Q,D$ and $b$ given by Equations~\eqref{eqn:source_general},\eqref{eqn:diffusion} and \eqref{eqn:energy_loss} above, the diffusion-loss equation can be solved for an equilibrium distribution of electrons. An example solution is provided in Figure~\ref{fig:coma_electrons}, which shows the full 2-dimensional electron equilibrium distribution that has been computed for the Coma galaxy cluster environment and a set of typical WIMP parameters. Note the larger concentration of low-energy electrons within the inner regions of the halo, which is a common feature of the equilibrium distributions. 

\subsection{Gas density and magnetic field}
A variety of gas density and magnetic field strength radial profiles are available by default within \verb|DarkMatters|. These consist of the constant (flat), power law (pl), $\beta$, double $\beta$, and exponential profiles. Every profile has a normalisation constant $n_0$ or $B_0$ for gas density and magnetic field respectively. We then use a notation system that a physical scale associated with the profile is either $r_e$ or $r_b$ while an exponent is $\beta_e$ or $\beta_b$. In the case of the double $\beta$ profile there exist two scales, normalisations, and exponents.  

Below we list the gas density and magnetic field strength profiles in their functional forms
\begin{align}\label{eqn:gas_profiles}
	n_{e,\text{flat}}(r) &= n_0	\,, \\
	n_{e,\text{pl}}(r) &= n_0 \left(\dfrac{r}{r_e}\right)^{\beta_e} \,, \\ 
	n_{e,\text{beta}}(r) &= n_0\left(1 + \left(\frac{r}{r_e}\right)^2 \right)^{\frac{3\beta_e}{2}} \,, \\
	n_{e,\text{d-beta}}(r) &= n_0\left(1 + \left(\frac{r}{r_e}\right)^2 \right)^{\frac{3\beta_e}{2}} + n_{0,2}\left(1 + \left(\frac{r}{r_{e,2}}\right)^2 \right)^{\frac{3\beta_{e,2}}{2}} \,, \\
	n_{e,\text{exp}}(r) &= n_0\exp\left(-\frac{r}{r_e}\right) \,,
\end{align}
and 
\begin{align}\label{eqn:magnetic_profiles}
	B_{\text{flat}}(r) &= B_0 \,,\\
	B_{\text{pl}}(r) &= B_0 \left(\dfrac{r}{r_b}\right)^{\beta_b} \,, \\
	B_{\text{beta}}(r) &= B_0\left(1 + \left(\frac{r}{r_b}\right)^2 \right)^{\frac{3\beta_b}{2}}\,, \\
	B_{\text{d-beta}}(r) &= B_0\left(1 + \left(\frac{r}{r_b}\right)^2 \right)^{\frac{3\beta_b}{2}} + B_{0,2}\left(1 + \left(\frac{r}{r_{b,2}}\right)^2 \right)^{\frac{3\beta_{b,2}}{2}}\,, \\
	B_{\text{exp}}(r) &= B_0\exp\left(-\frac{r}{r_b}\right)\,.
\end{align}

\subsection{Observables from DM}\label{sec:observables}

Once the physical environment of the DM halo and the particle model is fully specified, various multi-messenger observables can be calculated in \pkg{DarkMatters}. There are three broad categories that we focus on, which include radio emissions from synchrotron radiation, gamma rays produced from both prompt annihilation and secondary mechanisms, and prompt neutrino emmission. The details of the implementation of each of these emissions are described below.  

\subsubsection{Radio}\label{sec:radio}

\begin{figure}[htbp]
  \centering
  \begin{subfigure}[t]{0.5\linewidth}
    \centering
    \includegraphics[width=\linewidth]{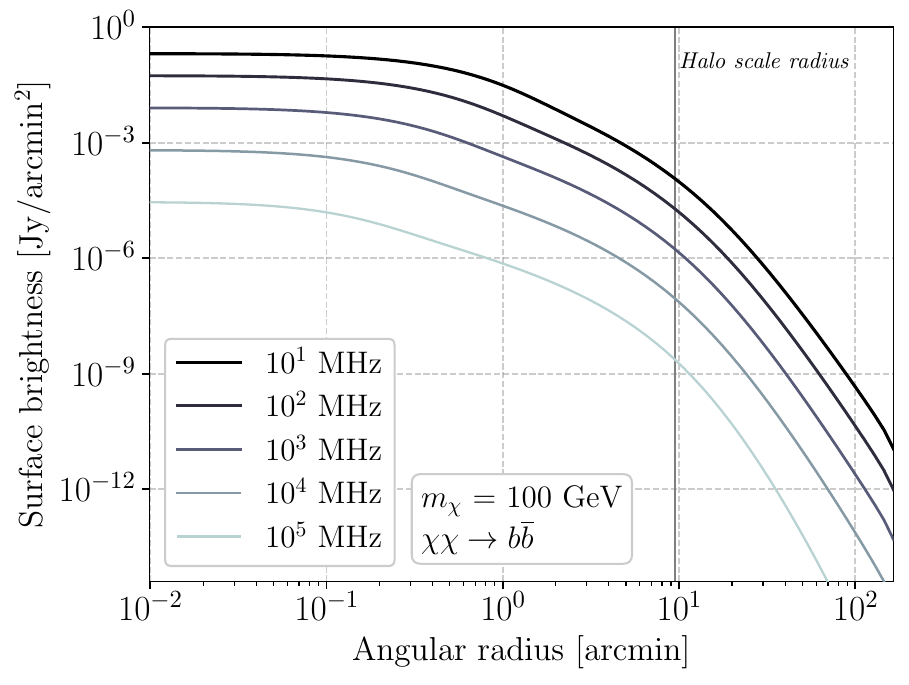}
    \caption{}
    \label{fig:coma_bb_sb}
  \end{subfigure}%
  \hfill%
  \begin{subfigure}[t]{0.5\linewidth}
    \centering
    \includegraphics[width=\linewidth]{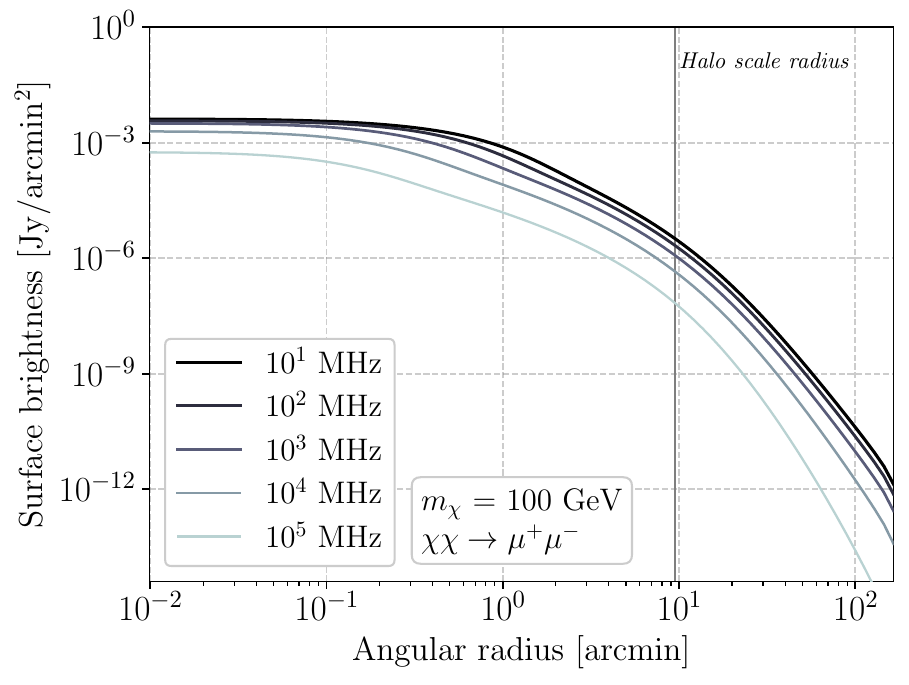}
    \caption{}    
    \label{fig:coma_mumu_sb}
  \end{subfigure}
  \caption{Sample calculations for the radio surface brightness emissions from the Coma galaxy cluster for a wide range of frequencies and two annihilation channels; (a): bottom quarks and (b): muons. Also shown is the characteristic scale radius $r_s$ of the DM halo, in equivalent angular units, as a vertical line. }
  \label{fig:coma_sbs}
\end{figure}

\begin{figure}[htbp]
  \centering
  \begin{subfigure}[t]{0.5\linewidth}
    \centering
    \includegraphics[width=\linewidth]{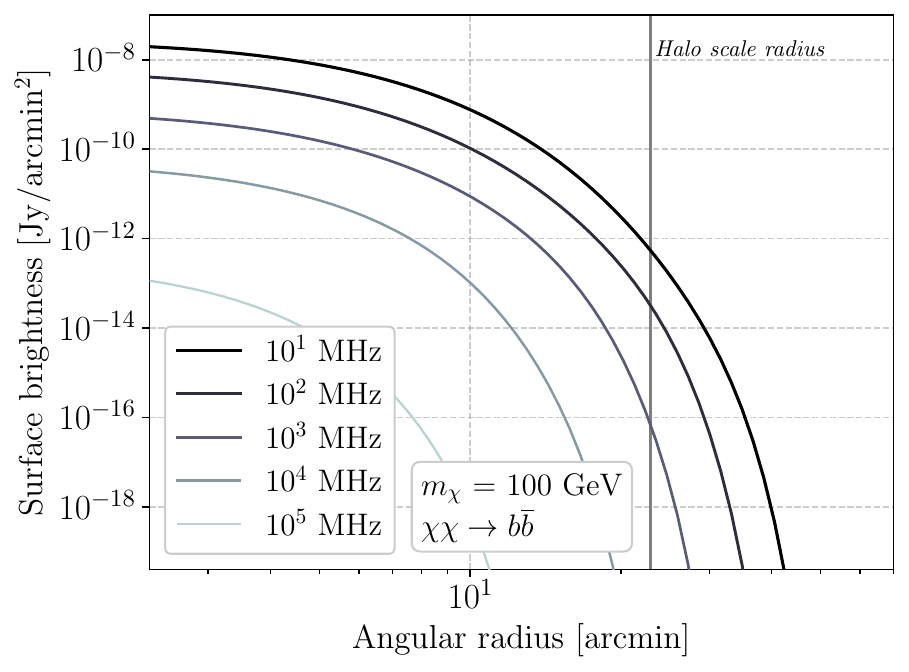}
    \caption{}
    \label{fig:retII_bb_sb}
  \end{subfigure}%
  \hfill%
  \begin{subfigure}[t]{0.5\linewidth}
    \centering
    \includegraphics[width=\linewidth]{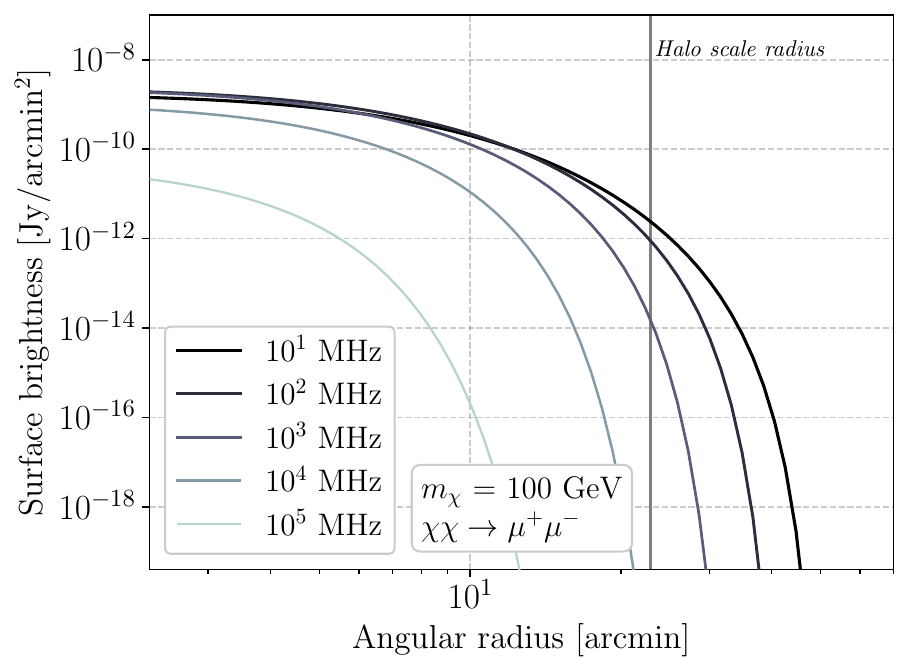}
    \caption{}    
    \label{fig:retII_mumu_sb}
  \end{subfigure}
  \caption{Equivalent surface brightness calculations as shown in Figure~\ref{fig:coma_sbs}, except for the Reticulum II dSph target.}
  \label{fig:retII_sbs}
\end{figure}

\begin{figure}[htbp]
  \centering
  \includegraphics[width=0.75\linewidth]{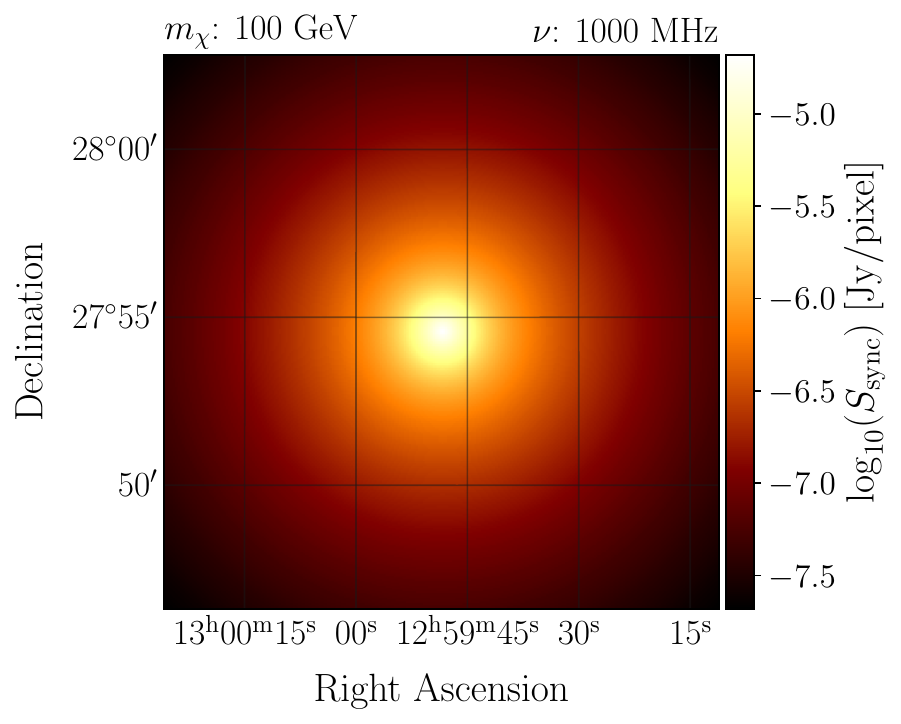}
  \caption{Sample 2-dimensional radio surface brightness image, computed by mapping the results of the radial surface brightness calculation in Equation~\eqref{eqn:surface_brightness} onto a 2-dimensional grid. Calculation parameters follow the sample for the Coma galaxy cluster with an NFW halo profile and 100 GeV WIMPs annihilating through the bottom quark channel. Results shown here are for synchrotron emission frequency of 1 GHz. Note that this image can be saved in the FITS data format by \pkg{DarkMatters}.}
  \label{fig:coma_bb_fits}
\end{figure}

The method used to calculate the radio emissivity from relativistic electrons follows a standard framework, as laid out in~\cite{colafrancescoMultifrequencyAnalysisNeutralino2006}. Firstly, the isotropically-averaged power of the synchrotron emission from a population of electrons is expressed as in~\cite{colafrancescoMultifrequencyAnalysisNeutralino2006}, 
\begin{equation}\label{eqn:synchrotron_power}
	P(\nu,E,r) = \int_0^{\pi} \dd{\theta} \frac{\sin(\theta)}{2} 2\pi \sqrt{3} \, r_e m_e c \nu_g \sin(\theta) F(x/\sin(\theta)) \,,
\end{equation}
where $r_e$, $m_e$ and $\nu_g$ are the electron's classical radius, mass and non-relativistic gyro-frequency, respectively. The function $F$ and the variable $x$ are defined as follows:
\begin{equation}\label{eqn:synchrotron_kernel}
	F(t) \equiv t\int^{\infty}_{t} \dd{z} K_{5/3}(z) 
\end{equation}
and
\begin{equation}
	x \equiv \frac{2\nu(1+z)}{3\nu_0\gamma^2}\left[ 1+\left( \frac{\gamma \nu_p}{\nu(1+z)} \right)^2 \right]^{3/2} \,,
\end{equation}
where $K_{5/3}$ is a modified Bessel function of order 5/3 and $\nu_p = 8980(n_e(r)/1 \,\mathrm{cm^{-3}})^{1/2}\;\mathrm{Hz}$ is the plasma frequency of the gas. We have used the fitting formulae provided by~\cite{foukaAnalyticalFitsSynchrotron2013} to calculate the kernel function in Equation~\ref{eqn:synchrotron_kernel}. The synchrotron power is then used to calculate the emissivity, by 
\begin{equation}\label{eqn:emissivity}
	j_{\mathrm{synch}}(\nu,r) = \int^{\mx}_{m_e} \dd E \left( \psi_-(r,E) + \psi_+(r,E) \right) P(\nu,E,r) \,,
\end{equation} 
where $\psi_{\pm}$ are the equilibrium electron and positron distributions found by solving Equation~\eqref{eqn:diffusion_loss}. Various common observable quantities, such as the radio surface brightness or the flux density spectrum, can now be calculated from the emissivity. Firstly, the radio surface brightness is defined as
\begin{equation}\label{eqn:surface_brightness}
	I_{\mathrm{synch}}(\nu,s) = \int \dd{l} \, \frac{j(\nu,\sqrt{s^2+l^2})}{4\pi}\,,
\end{equation}
where $l$ is the line-of-sight coordinate along a cylinderical axis and $s$ is the corresponding cylindrical radius to a point within the halo. Sample results calculated using Equation~\eqref{eqn:surface_brightness} can be seen in Figures~\ref{fig:coma_sbs} and~\ref{fig:retII_sbs}. As modern radio datasets are often accessible as 2-dimensional images, \pkg{DarkMatters} provides a mapping of the above surface brightness profile onto a 2-dimensional spatial grid, which allows for convenient comparisons between model and data. The sky coordinates and desired pixel size or image resolution of the produced image are all adjustable parameters when interfacing with the code, and \pkg{DarkMatters} also provides an option to save these images conformant to the FITS data file format~\footnote{\url{https://fits.gsfc.nasa.gov/}}, which is a standard multi-dimensional data format used extensively in the radio astronomy community. A sample 2-dimensional image of radio surface brightness emissions (calculated with Equation~\eqref{eqn:surface_brightness}) is shown in Figure~\ref{fig:coma_bb_fits}.

\begin{figure}[htbp]
  \centering
    \begin{subfigure}[t]{0.5\linewidth}
      \centering
      \includegraphics[width=\linewidth]{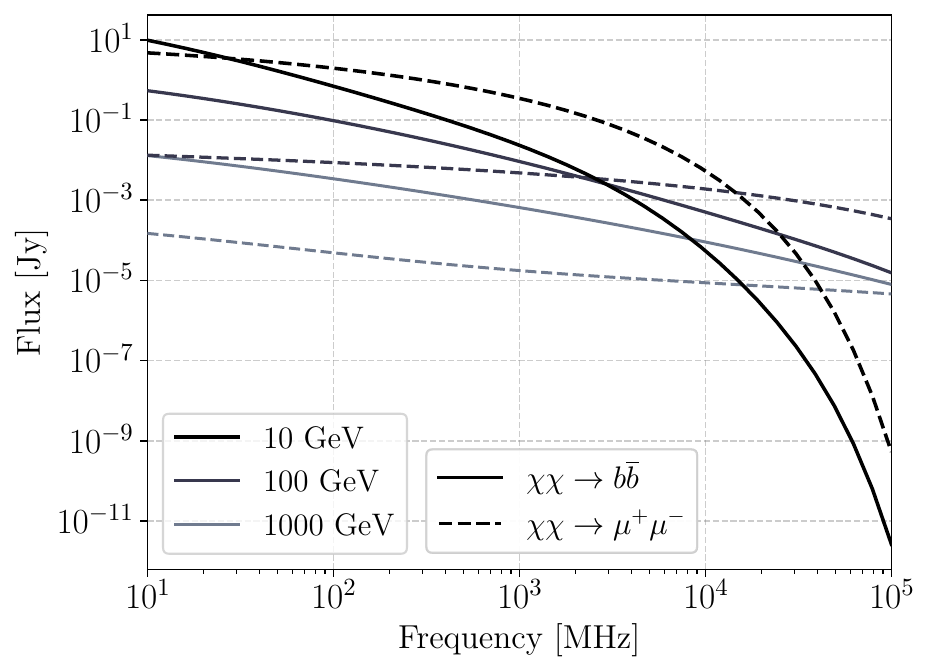}
      \caption{}
      \label{fig:coma_flux}
    \end{subfigure}%
  \hfill%
  \begin{subfigure}[t]{0.5\linewidth}
    \centering
    \includegraphics[width=\linewidth]{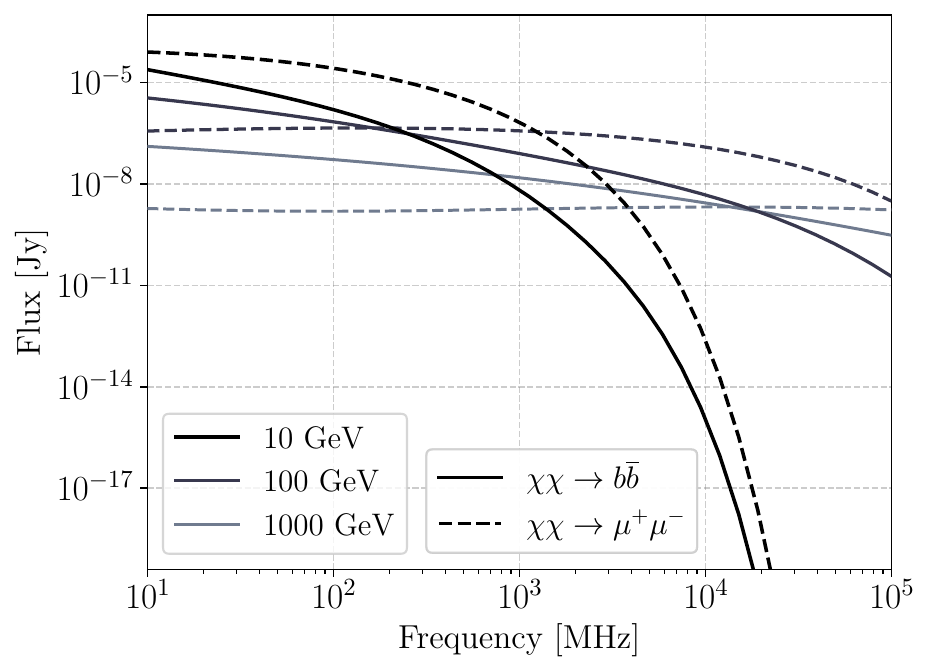}
    \label{fig:retII_flux}
    \caption{}
  \end{subfigure}
  \caption{Sample calculations for the integrated radio flux density from the (a): Coma galaxy cluster and (b): Reticulum II dSph. Results shown vary the WIMP mass and annihilation channel for comparative purposes. Note the change in flux scale between the two figures.}
  \label{fig:fluxes}
\end{figure}

The final radio-frequency observable that we consider is the integrated radio flux density, defined as
\begin{equation}\label{eqn:integrated_flux}
	S_{\mathrm{synch}}(\nu,R) = \int_0^R \dd[3]{r'} \frac{j(\nu,r')}{4\pi\left(d_L^2+(r')^2\right)}\,,
\end{equation}
where $R$ represents the maximum radius of the halo within which the emission is integrated, and $d_L$ is the luminosity distance to the halo. A set of sample calculations of the integrated flux, as calculated with Equation~\eqref{eqn:integrated_flux}, is shown in Figure~\ref{fig:fluxes}.

\subsubsection{Gamma-ray}\label{sec:gamma-ray}

\begin{description}
  \item[Primary (prompt) $\gamma$'s] The differential flux of gamma rays that are produced directly from WIMP annihilations and decay, referred to as the `prompt' emission, is calculated as follows. 
  
  \begin{equation}\label{eqn:j_factor}
    J(\Delta\Omega,l) = \int_{\Delta\Omega}\int_{l} \rho^2(\mathbf{r}^{\prime}) \dd l^{\prime} \dd \Omega^{\prime} \,, 
  \end{equation}
  
  \begin{equation}\label{eqn:d_factor}
    D(\Delta\Omega,l) = \int_{\Delta\Omega}\int_{l} \rho(\mathbf{r}^{\prime}) \dd l^{\prime} \dd \Omega^{\prime} \,, 
  \end{equation}
  
  \begin{equation}\label{eqn:integrate_flux_gamma}
    S_{\gamma}(E) \equiv \dv{\Phi_{\gamma}}{E} = 
    \begin{cases}
      \displaystyle \frac{\cs}{8\pi\mx^2} \, J \, \dv{N_i}{E} \,, &\qq{(annihilation)} \\[1.0em]
      \displaystyle \frac{\Gamma}{4\pi\mx} \, D \, \dv{N_i}{E} \,, &\qq{(decay)} 
  \end{cases} 
\end{equation}

where the energy spectra of the produced gamma rays are found from Equation~\eqref{eqn:particle_spectrum}.\\

\item[Secondary (ICS) $\gamma$'s] This is composed of high-energy photons produced via the equilibrium electron population within the DM halo. The mechanisms used are Inverse Compton Scattering (ICS) and bremsstrahlung. 

The ICS power at observed frequency $\nu$ from an electron with energy $E$ at redshift $z$ is determined via~\cite{longairHighEnergyAstrophysics2011,rybicki1986}
\begin{equation}
P_{IC} (\nu,E,z) = c E_{\gamma}(z) \int d\epsilon \; n(\epsilon) \sigma(E,\epsilon,E_{\gamma}(z)) \; ,
\label{eq:ics_power}
\end{equation}
where $E_\gamma (z) = h \nu (1+z)$, $\epsilon$ is the seed photon energy distributed according to $n(\epsilon)$. The Klein-Nishina cross-section is defined according to
\begin{equation}
\sigma(E,\epsilon,E_{\gamma}) = \frac{3\sigma_T}{4\epsilon\gamma^2}G(q,\Gamma_e) \; ,
\end{equation}
where $\sigma_T$ is the Thompson cross-section and
\begin{equation}
G(q,\Gamma_e) = 2 q \ln{q} + (1+2 q)(1-q) + \frac{(\Gamma_e q)^2(1-q)}{2(1+\Gamma_e q)} \; ,
\end{equation}
with
\begin{equation}
\begin{aligned}
q & = \frac{E_{\gamma}}{\Gamma_e(\gamma m_e c^2 + E_{\gamma})} \; , \\[0.5em]
\Gamma_e & = \frac{4\epsilon\gamma}{m_e c^2} \; ,
\end{aligned}
\end{equation}
where $\gamma$ is the electron Lorentz factor. 

For bremsstrahlung of an electron colliding with target species $j$ we use~\cite{longairHighEnergyAstrophysics2011,rybicki1986}
\begin{equation}
P_B (\nu,E,r) = c E_{\gamma}(z)\sum\limits_{j} n_j(r) \sigma_B (E_{\gamma},E) \; ,
\end{equation}
where the notation remains the same as for ICS. Note that $n_j$ is the target species number density and the cross-section is given by
\begin{equation}
\sigma_B (E_{\gamma},E) = \frac{3\alpha \sigma_T}{8\pi E_{\gamma}}\left[ \left(1+\left(1-\frac{E_{\gamma}}{E}\right)^2\right)\phi_1 - \frac{2}{3}\left(1-\frac{E_{\gamma}}{E}\right)\phi_2 \right] \; ,
\end{equation}
where $\phi_1$ and $\phi_2$ are species dependent factors~\cite{longairHighEnergyAstrophysics2011,rybicki1986}. In \verb|DarkMatters| we use only protons as the target species and equate $n_j$ to the gas density within the halo environment.

The fluxes from the mechanisms is determined following the same process as for radio emissions, but with the appropriate choice of power function in Eq.~\ref{eqn:emissivity}.

\end{description}

\begin{figure}[htbp]
  \centering
    \begin{subfigure}[t]{0.5\linewidth}
      \centering
      \includegraphics[width=\linewidth]{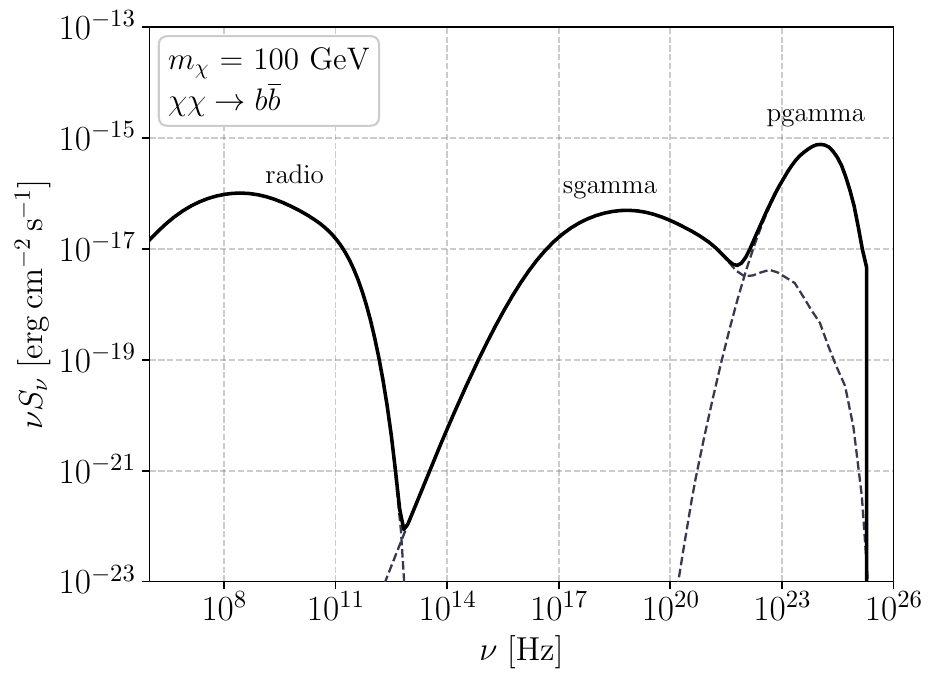}
      \caption{}
      \label{fig:coma_sed}
    \end{subfigure}%
  \hfill%
  \begin{subfigure}[t]{0.5\linewidth}
    \centering
    \includegraphics[width=\linewidth]{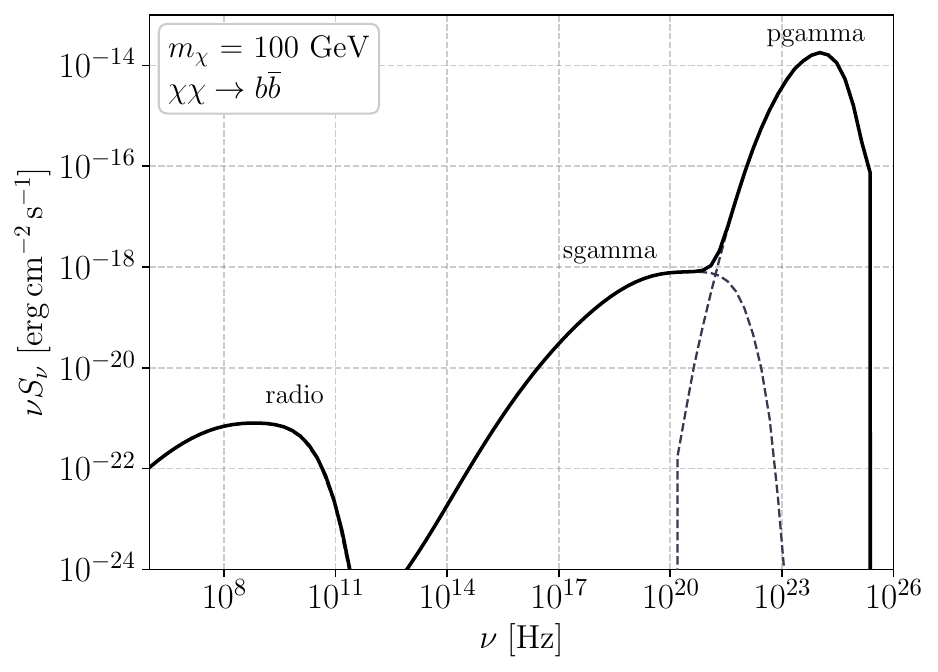}
    \label{fig:retII_sed}
    \caption{}
  \end{subfigure}
  \caption{Spectral Energy Distribution (SED) for the multi-wavelength WIMP-related emissions from the (a): Coma galaxy cluster and (b): Reticulum II dSph. The dominant contributions from each major emission type (corresponding to the radio synchrotron, prompt gamma-ray and secondary gamma-ray emissions described above) are labelled at the relevant locations on the curves.}
  \label{fig:seds}
\end{figure}

\subsubsection{Neutrino}\label{sec:neutrino}
\begin{figure}[htbp]
  \centering
  \includegraphics[width=0.75\linewidth]{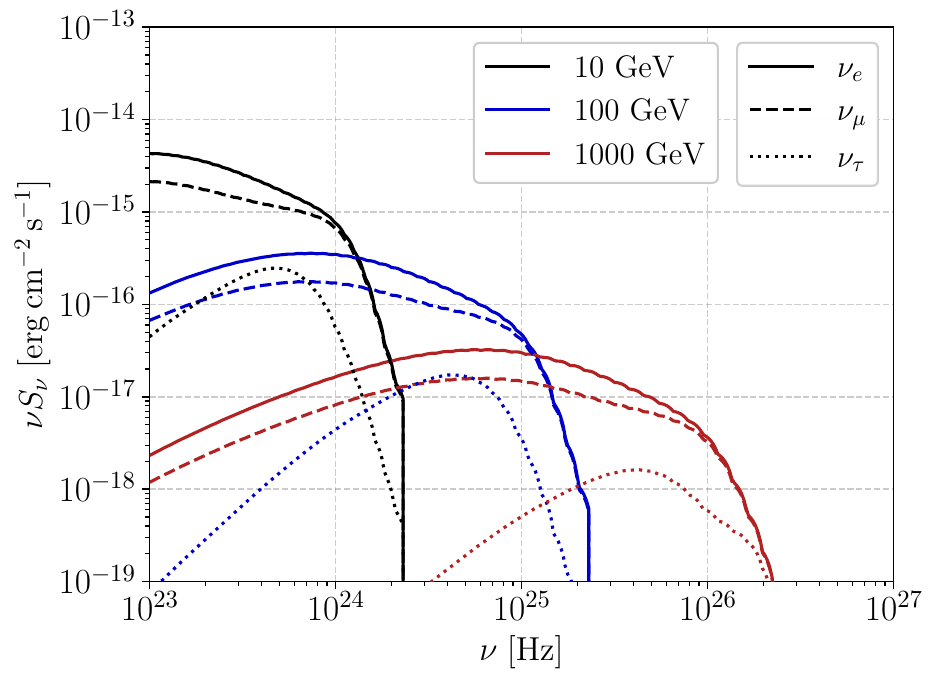}
  \caption{Sample neutrino flux output from the Coma galaxy cluster, calculated using Equation~\eqref{eqn:integrate_flux_neutrino}.}
  \label{fig:coma_neutrinos}
\end{figure}

Neutrino fluxes are computed in the same manner as prompt gamma-rays
\begin{equation}\label{eqn:integrate_flux_neutrino}
    S_{\nu}(E) \equiv \dv{\Phi_{\nu}}{E} = 
    \begin{cases}
      \displaystyle \frac{\cs}{8\pi\mx^2} \, J \, \dv{N_\nu}{E} \,, &\qq{(annihilation)} \\[1.0em]
      \displaystyle \frac{\Gamma}{4\pi\mx} \, D \, \dv{N_\nu}{E} \,, &\qq{(decay)} 
  \end{cases} 
\end{equation}
where $J$ and $D$ have been previously defined in Eqs.~(\ref{eqn:j_factor}) and (\ref{eqn:d_factor}). A set of sample calculations using the above equation can be seen in Figure~\ref{fig:coma_neutrinos} for the Coma galaxy cluster.

\section{Implementation and usage details}\label{sec:implementation}

\subsection{Installation}
The package is installed merely by cloning the GitHub repository and ensuring all the dependencies are installed (a file \verb|requirements.txt| is provided for this purpose). 

\subsection{Specifying input files}\label{sec:input_files}
\verb|DarkMatters| uses either yaml or json files for input and output, the default option is yaml. Below we specify a yaml input for an example calculation

\begin{verbatim}
halo_data:
  name: "segue1"
  profile: "einasto"
  index: 0.3333
  rho_norm: 
    value: 1.738e+8
    unit: "Msun/kpc^3"
  scale: 
    value: 0.15
    unit: "kpc"
  distance: 
    value: 23.0
    unit: "kpc" 

mag_data:
  profile: "flat"
  mag_norm: 
    value: 2.e-6
    unit: "gauss"

gas_data:
  profile: "exp"
  gas_norm: 
    value: 1.e-4
    unit: "1/cm^3"
  scale: 
    value: 38.0
    unit: "pc" 
 
diff_data:
  diff_index: 0.7
  diff_rmax:
    value: 1.6
    unit: "kpc"
  diff_constant:
    value: 3.0e+26
    unit: "cm^2/s"

part_data:
  part_model: "bb"
  em_model: "annihilation"

calc_data:
  calc_mode: "flux"
  freq_mode: "radio"
  m_wimp:
    value:
      - 10
      - 100
      - 1000
    unit: "GeV"
  f_sample_limits:
    value:
      - 5.e+1
      - 2.e+4
    unit: "MHz"
  f_sample_num: 40
  f_sample_spacing: "log"
  calc_angmax_integrate:
    value: 4
    unit: "degree"
  os_delta_t_min:
    value: 0.1
    unit: "yr"
\end{verbatim}

The above example input also serves to illustrate the data structures used in the package, these being a set of python dictionaries \verb|halo_data|, \verb|mag_data|, \verb|gas_data|, \verb|diff_data|, \verb|part_data|, and \verb|calc_data|. These store the properties associated with the halo, magnetic field, gas, diffusion environment, particle physics, and calculation to be performed. Another feature to remark on is the specification of unitful quantities: they are given by a dictionary with a `value' and `unit' entry. Any unit of same type (mass,distance,density) can be given and \verb|astropy| is used to convert it into the internal unit system of \verb|DarkMatters|. This requires that the unit entry be specified in an \verb|astropy| friendly format. 

One other dictionary exists: \verb|cosmo_data|. This contains the cosmological parameters to be used in calculations. If left unspecified, it defaults to the Planck 2018 results~\cite{planckcollaborationPlanck2018Results2020}.

\subsubsection{halo\_data}
Here we specify the information on the DM halo. A summary of how modelling parameters link to the dictionary is presented in Table~\ref{tab:halo-profiles}. Note that the density profiles can be accessed by specifying the \verb|halo_profile| parameter with the allowed values `nfw', `gnfw', `einasto', `burkert', and `isothermal'. Additional profiles can be added by modifying the \texttt{halo\_density\_profiles.yaml} file, in the \verb|dark_matters/config| folder, as well as adding a function recipe to the \verb|halo_density_builder| function in the \linebreak\verb|dark_matters/astro_cosmo/astrophysics.py| file. 

\begin{table}[ht!]
    \centering
    \caption{Examples of \texttt{DarkMatters} halo parameter correspondence to modelling.}
    \label{tab:halo-profiles}
    \begin{tabular}{llll}
    \hline
    Quantity & Variable & Type & Unit \\
    \hline
    $M_{\mathrm{vir}}$ & \verb|mvir| & Float & Mass\\
    $R_{\mathrm{vir}}$ & \verb|rvir| & Float & Distance\\
    $c_{\mathrm{vir}}$ & \verb|cvir| & Float & -\\
    $r_s$ & \verb|scale| & Float & Distance\\
    $\rho_s$ & \verb|rho_norm| & Float & Mass density\\
    $\frac{\rho_s}{\rho_c}$ & \verb|rho_norm_relative|  & Float & - \\
    $\alpha$ & \verb|index| & Float & -\\
    $\rho(r)$ & \verb|profile| & String & - \\
    $d_L$ & \verb|distance| & Float & Distance \\
    $z$ & \verb|z| & Float & - \\
    \hline
    \end{tabular}
\end{table}

\subsubsection{mag\_data}
The magnetic field data is linked to the modelling parameters in Table~\ref{tab:mag}. The common options for \verb|mag_profile| are `flat', `powerlaw', `beta', `doublebeta', and `exp'.
\begin{table}[ht!]
    \centering
    \caption{Examples of \texttt{DarkMatters} magnetic field parameter correspondence to modelling from Eq.~(\ref{eqn:magnetic_profiles}).}
    \label{tab:mag}
    \begin{tabular}{lllll}
    \hline
    Quantity & Variable & Type & Unit \\
    \hline
    $B_0$ & \verb|mag_norm| & Float & Magnetic field strength\\
    $\beta_b$ & \verb|index| & Float & -\\
    $r_b$ & \verb|scale| & Float & Distance\\
    $B(r)$ & \verb|profile| & String & - \\
    \hline
    \end{tabular}
\end{table}

\subsubsection{gas\_data}
The gas density data is linked to the modelling parameters in Table~\ref{tab:gas}. The common options for \verb|gas_profile| are `flat', `powerlaw', `beta', 'doublebeta', and `exp'.
\begin{table}[ht!]
    \centering
    \caption{Examples of \texttt{DarkMatters} gas density parameter correspondence to modelling from Eq.~(\ref{eqn:gas_profiles}).}
    \label{tab:gas}
    \begin{tabular}{lllll}
    \hline
    Quantity & Variable & Type & Unit \\
    \hline
    $n_0$ & \verb|gas_norm| & Float & Number density\\
    $\beta_e$ & \verb|index| & Float & -\\
    $r_e$ & \verb|scale| & Float & Distance\\
    $n(r)$ & \verb|profile| & String & - \\
    \hline
    \end{tabular}
\end{table}

\subsubsection{diff\_data}
This dictionary is quite simple, containing the radius at which the boundary conditions are applied \verb|diff_rmax| as well as $D_0$, the diffusion constant, in \verb|diff_constant|. The parameter \verb|diff_index| specifies the index $\delta$ in Eq.~(\ref{eqn:diffusion}).

It is also possible to choose the target photon field for ICS by specifying the photon temperature (at present the code only accommodates black-body target fields) through the parameter \verb|photon_temp|. Additionally, the energy density of ICS seed photons (for calculating energy loss in Eq.~(\ref{eqn:energy_loss})) can be specified via \verb|photon_density| which has units of energy density. 

\subsubsection{part\_data}
Here we specify the particle physics data. The annihilation/decay channel is given by \verb|part_model| and we choose between `annihilation' or `decay' options using \verb|em_model|. Custom models can also be included by setting up an input file with appropriate structure. See the Wiki~\footnote{\url{https://github.com/Hyperthetical/DarkMatters/wiki}} for more detail. 

\subsubsection{cosmo\_data}
This is not displayed as it defaults to Planck 2018~\cite{planckcollaborationPlanck2018Results2020}. Note the numerical fitting function for $c_\mathrm{vir}$ can be chosen here using the parameter \verb|cvir_mode|. The options are `p12'~\cite{pradaHaloConcentrationsStandard2012}, `munoz\_2011'~\cite{munoz2011}, `cpu\_2006'~\cite{colafrancescoMultifrequencyAnalysisNeutralino2006} and `bullock\_2001'~\cite{bullockProfilesDarkHaloes2001}. Note that the default value is `p12'. 

\subsubsection{calc\_data}
In this dictionary we actually specify what we want to calculate with all the other input data. This is principally chosen via two parameters: \verb|calc_mode| and \verb|freq_mode|. The first can take options `flux', `sb', or `jflux'. These refer to integrated flux, surface brightness, and flux from a J or D factor respectively.  The second parameter decides on the mechanisms to be considered. Its options are `all', `gamma', `radio', `pgamma', `sgamma' or `neutrinos\_x' (where x can be mu, tau, or e). These refer to: all electromagnetic emissions, primary plus secondary high-energy emissions, radio only, primary high-energy only, secondary high-energy only, and neutrinos respectively.

Further parameters specify more detail. For example \verb|m_wimp| provides a list of WIMP masses (in energy units) to perform the calculation for. Then \verb|f_sample_limits|, \verb|f_sample_num|, and \verb|f_sample_spacing| decide what frequencies to perform the calculation at. These can be manually specified using \verb|f_sample_values|, which provides a list of the values, instead. The parameter \verb|calc_angmax_integrate| gives the radius within which to integrate the flux (a physical radius can be specified instead with \verb|calc_rmax_integrate|). Finally, \verb|os_delta_t_min| tells the algorithm the smallest time spacing value to use. This tends to need to be adjusted from the default only for small structures like dwarf galaxies (see the Wiki for further detail).

\subsection{Running the code}
The following simple script will run \verb|DarkMatters| 
\begin{verbatim}
    import sys
    sys.path.append(dark_matters_path)
    from dark_matters.input import read_input_file
    from dark_matters.calculations import run_calculation
    from dark_matters.output import make_output

    data_sets = read_input_file("example.yaml")
    output_data = run_calculation(**data_sets)
    make_output(**output_data,out_mode="yaml")
\end{verbatim}
This assumes the path to the folder containing the \verb|dark_matters| folder is specified in place of \verb|dark_matters_path| and that the input file is called \verb|example.yaml|. The code will produce an output yaml file with a name that reflects some of the associated calculation parameters. This file will contain all of the input dictionaries and include the results in \verb|calc_data["results"]|, which is itself a dictionary. The file contents is identical to the contents of the \verb|output_data| variable above.

When the code is executed it conducts a set of consistency checks on the input dictionaries and calculates all the supplementary parameters needed to perform the requested tasks. It then proceeds to execute the instructions in the \verb|calc_data| dictionary. 

\section{Comparisons with existing tools}\label{sec:comparisons}

Aside from an internal test suite, a set of simple calculations with the Radio and X-ray DMFIT (\prx{}) package~\cite{mcdanielMultiwavelengthAnalysisDark2017} have been performed to compare and validate the results from \pdm{}. The \prx{} package\footnote{\url{https://github.com/alex-mcdaniel/RX-DMFIT}} was originally written to extend the \pkg{DMFIT}~\cite{jeltemaFittingGammaRaySpectrum2008} package with radio and X-ray fluxes. At the time of writing, surface brightness outputs (of the form in Equation~\eqref{eqn:surface_brightness}) are not supported in \prx{}, so the comparisons are focused on emissivity and integrated flux outputs. For this, we have considered two typical and common targets of study in the literature -- the Coma galaxy cluster and the Reticulum II dSph. These targets exist on opposite ends of size and mass ranges for astrophysical structures, and provide edge cases for the physical environments that our package is designed to process. For these tests, we have kept the input configurations between the two codes as similar as possible, to highlight differences in the underlying method rather than any differences in parameter choices. The input configurations used in these tests have been provided as online material alongside this manuscript through the Zenodo doi\footnote{\url{https://doi.org/10.5281/zenodo.13312389}.}

\begin{figure}[htbp]
  \centering
  \begin{subfigure}[t]{0.5\linewidth}
    \centering
    \includegraphics[width=\linewidth]{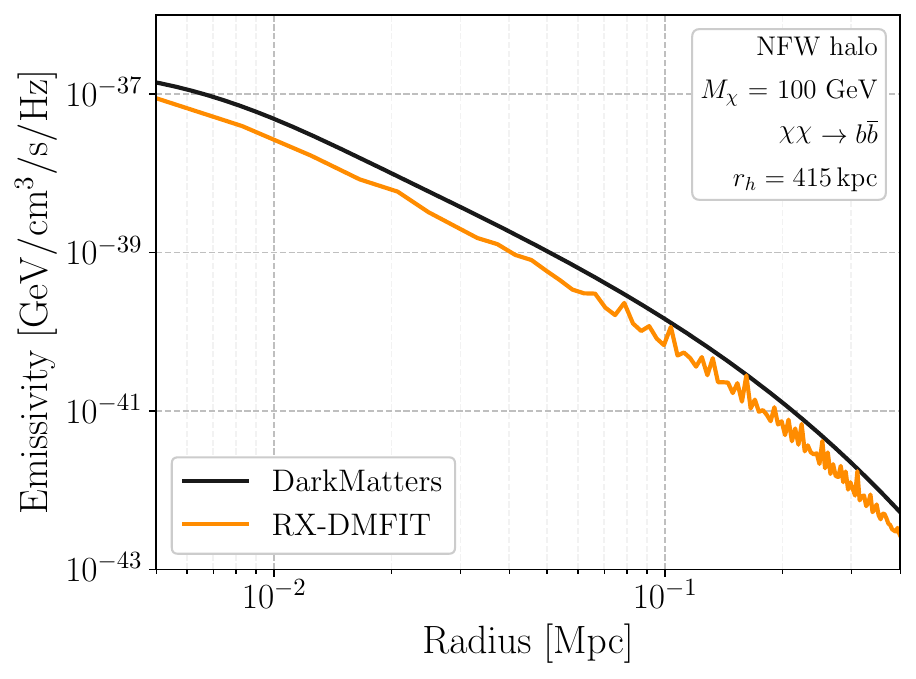}
  \end{subfigure}%
  \hfill%
  \begin{subfigure}[t]{0.5\linewidth}
    \centering
    \includegraphics[width=\linewidth]{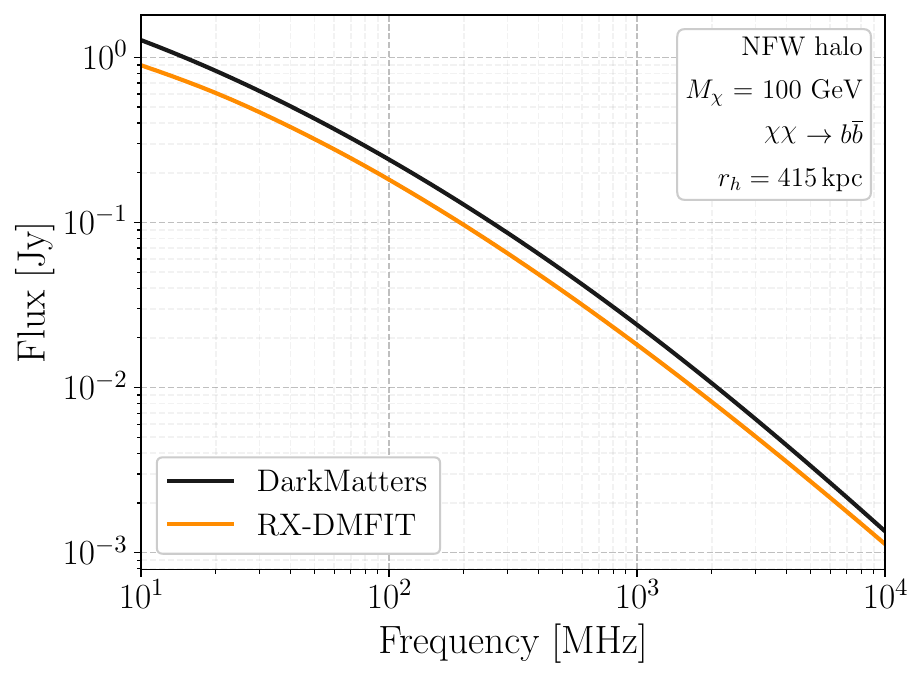}
  \end{subfigure}
  \caption{Direct comparison of outputs calculated using the \pdm{} (black curve) and \prx{} (orange curve) software packages. Results shown here are the emissivity (Equation~\eqref{eqn:emissivity}) in the left panel and the integrated flux (Equation~\eqref{eqn:integrated_flux}) in the right panel, for the Coma galaxy cluster modelled with parameters shown on the figure. For a full list of parameters used in these calculations, see the supplementary online material provided with this manuscript.}
  \label{fig:coma_RXDMFIT}
\end{figure}

\begin{figure}[htbp]
  \centering
  \begin{subfigure}[t]{0.5\linewidth}
    \centering
    \includegraphics[width=\linewidth]{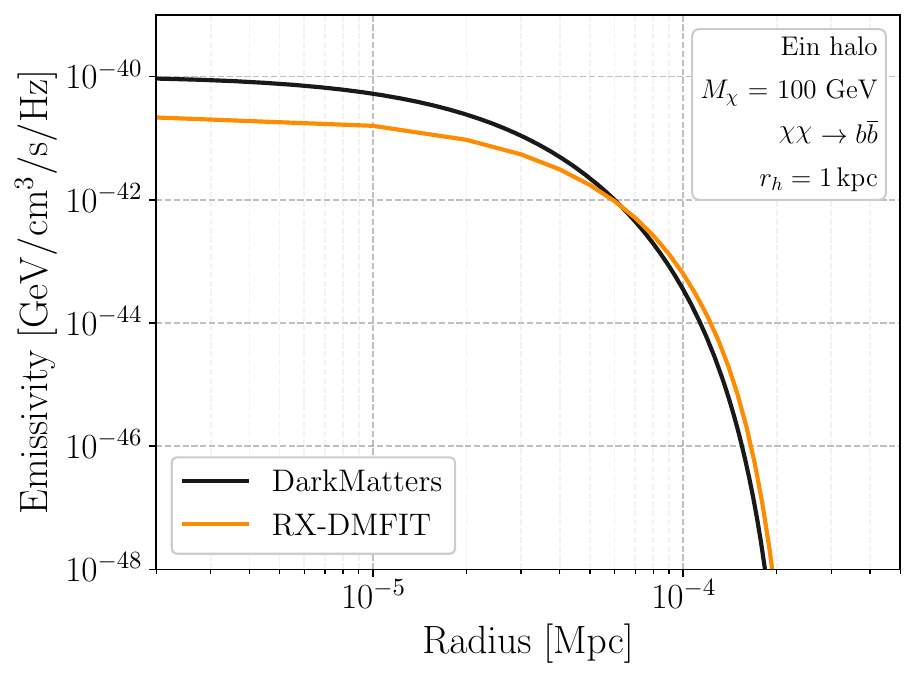}
  \end{subfigure}%
  \hfill%
  \begin{subfigure}[t]{0.5\linewidth}
    \centering
    \includegraphics[width=\linewidth]{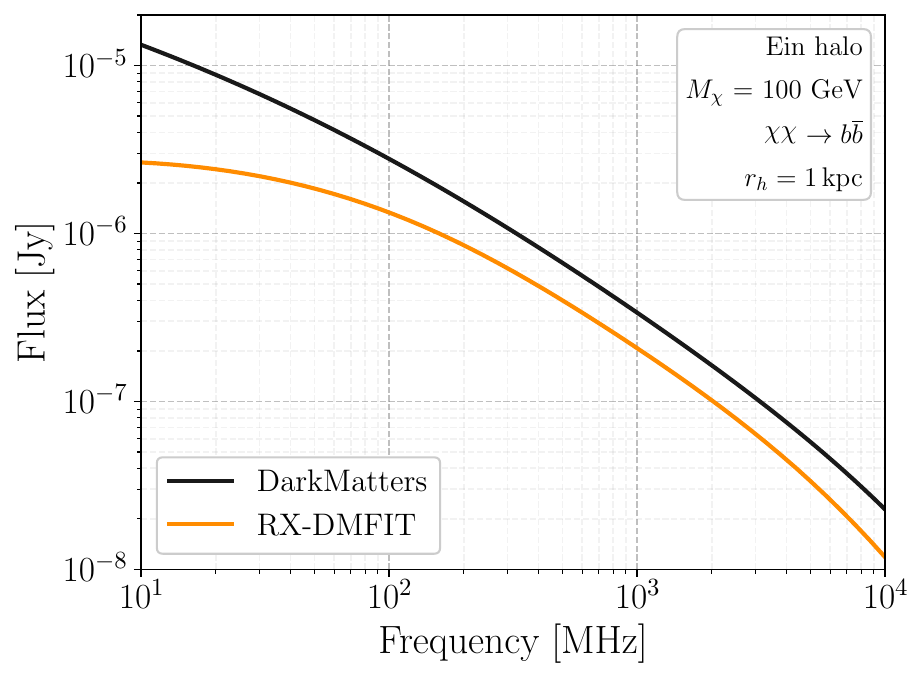}
  \end{subfigure}
  \caption{Direct comparison of outputs calculated using the \pdm{} (black curve) and \prx{} (orange curve) software packages. Results shown here are the emissivity (Equation~\eqref{eqn:emissivity}) in the left panel and the integrated flux (Equation~\eqref{eqn:integrated_flux}) in the right panel, for the Reticulum~II dSph modelled with parameters shown on the figure. For a full list of parameters used in these calculations, see the supplementary online material provided with this manuscript.}
  \label{fig:retII_RXDMFIT}
\end{figure}

The comparisons between both the radio emissivity and integrated flux are displayed in Figures~\ref{fig:coma_RXDMFIT} and~\ref{fig:retII_RXDMFIT}. We have also calculated simple relative differences between the methods to quantify the changes, where each result has been found relative to the larger value, i.e. we take $(\abs{y_1-y_2}/y_1)\times 100\%$, where $y_1 > y_2$. In the Coma galaxy cluster, we see \pdm{} produce $\sim46\%$ higher emissivity values throughout the halo. The integrated flux values are likewise higher, by $\sim30\%$ at 10~MHz, down to $\sim15\%$ at 10~GHz with an average of $\sim23\%$ difference over all frequencies in this range. In Reticulum~II, we observe larger overall differences, with the emissivities from \pdm{} higher by $\sim76\%$ at the center of the halo, dropping to $\sim60\%$ when averaged over all radii. Notably, the emissivities converge at a point within the halo, with values from \prx{} larger at large radii -- this trend will be discussed further below. The integrated flux outputs from Reticulum~II have a maximum difference of $\sim80\%$ at 10~MHz, which drops to $\sim50\%$ averaged over all frequencies.

The trends between each package's results shown above can be attributed to the treatment of the spatially dependent variables in the solution to the diffusion-loss equation. This is one of the primary differences between the two packages, with \prx{} using the semi-analytical technique first described in~\cite{colafrancescoMultifrequencyAnalysisNeutralino2006}, and \pdm{} using the numerical technique outlined in~\ref{sec:solution_methods}. One consequence of using the semi-analytical method of~\cite{colafrancescoMultifrequencyAnalysisNeutralino2006} is that all parameters involved in the diffusion-loss equation cannot depend on spatial variables (in this case the radius variable $r$). To illustrate how this point affects the observable emissions, consider first that a factor of the magnetic field strength appears in both the diffusion and energy-loss functions (Equations~\eqref{eqn:diffusion} and~\eqref{eqn:energy_loss} respectively). To satisfy the above condition, an average value for the halo is calculated in \prx{} according to 
\begin{equation}\label{eqn:rxdmfit_mag}
    B_{\text{avg}} = \frac{1}{r_h}\int_0^{r_h} \dd{r} B(r)\,,
\end{equation}
where $r_h$ defines the maximum radius of the diffusive environment~\cite{mcdanielMultiwavelengthAnalysisDark2017}. As the relevant astrophysical magnetic field profiles $B(r)$ monotonically decrease from their central strength $B_0$, in general we have $B_{\text{avg}} < B_0$, especially when $r_h$ is larger than the scale radius of the magnetic field. Since synchrotron radiation (which depends on $B(r)$) is the dominant energy-loss mechanism at higher electron energies, the overall energy-loss function $b(E)$ would be reduced at the center of the halo when an average value of the magnetic field is used over the full profile. Now, the semi-analytical solution method of~\cite{colafrancescoMultifrequencyAnalysisNeutralino2006,mcdanielMultiwavelengthAnalysisDark2017} makes use of a set of Green's functions that are defined with a variable $v(E)$ as follows:
\begin{equation}\label{eqn:diffusion_scale}
    v(E) = \int^{M_{\chi}}_{E} \dd{\tE} \frac{D(\tE)}{b(\tE)} \, .
\end{equation}
The quantity $\sqrt{v}$, which represents the mean distance that an electron would travel while losing a specified amount of energy, depends inversely on $b(E)$. Lower values of $b(E)$ that are caused by averaging the magnetic field will thus result in larger values of $v$ and electrons that diffuse further in the halo while losing the same energy. 

This interpretation can be used to explain the differences in the emissivity results seen between the two packages. With a central magnetic field strength that is relatively large in the \pdm{} case, the electrons are confined to the central regions of the halo where they continue to radiate. However, with the larger diffusion scale length determined by the lower overall magnetic field strength, electrons in the \prx{} case travel further through the halo to the outer regions, while losing an equivalent amount of energy. This effect can be seen in Figure~\ref{fig:retII_RXDMFIT}, which shows relatively less emissivity from \prx{} at the center of the halo, but slightly more as $r\rightarrow r_h$. This is likely a stronger effect in Reticulum~II than in Coma (Figure~\ref{fig:coma_RXDMFIT}) because of the steeper exponential magnetic field profile found and comparatively smaller diffusion region in Reticulum~II, both of which more effectively reduce the average magnetic field strength from the central value of $B_0$. Then, we note that the integrated flux calculations show larger fluxes from \pdm{} for all frequencies. This is expected behaviour given the larger emissivity values for each target, especially at the center of the halo where the emission is strongest. 

Finally, we note the presence of some minor numerical artefacts in the emissivity calculations from \prx{} at large radii in the Coma cluster, which are also present in the results displayed in~\cite{mcdanielMultiwavelengthAnalysisDark2017}. These artefacts have been observed to worsen significantly with different values of the diffusion radius $r_h$, especially if this value becomes comparable to or larger than the virial radius of the DM halo. The artefacts seem to originate from the numerical integration scheme used in the solution method of~\cite{colafrancescoMultifrequencyAnalysisNeutralino2006}, and should be accounted for in the \prx{} package through the use of alternative integration schemes provided by the GNU Scientific Library (GSL)~\cite{galassiGNUScientificLibrary2009}. We note that the results shown here were found with the default method, a quadrature non-adaptive Gauss-Kronod technique (QNG), though the quadrature adaptive integration with singularities (QAGS) method produced similar results. The use of the recommended scheme options for computational speed in \prx{}, i.e. the use of QNG integration, with a lookup table for the Green's function calculations, resulted in a average computation time of 295 seconds (taken over all performed calculations on the author's personal desktop computer). In comparison, the average computation time for the results from \pdm{}, performed on the same hardware, was 28 seconds. Of this time, an average of 24 seconds was used for finding the equilibrium distribution of electrons as the solution to the diffusion-loss equation, with the remaining time taken for emissivity or flux calculations. Profiling the code reveals that the vast majority of the compute time is made up of calls to the linear algebra method used to solve the matrix equations described in~\ref{sec:solution_methods}, and the specifics of these methods will be discussed further therein. The code profiling results shown here correspond to the claim made in~\cite{regisLocalGroupDSph2015a} that the numerical solution method is faster than the semi-analytical one. The use of advanced numerical techniques -- like completely vectorised, sparse, block-matrix algebra with dedicated \texttt{SciPy}~\cite{scipy2020} libraries that are designed for fast numerical calculations and an accelerated timestep-switching technique inspired by \pkg{GALPROP}, have further improved computational performance of the \pdm{} package. 

\section{Discussion and conclusions}\label{sec:conclusion}

In this paper we have introduced \pdm{}, the codebase which was developed to calculate all of the relevant physical properties of WIMPy DM halos and simulate their observable multiwavelength emissions. The functionality and configurability of the software should provide a complete solution -- from basic input parameters of the system to final observable quantities -- for a wide range of astrophysical environments. To fulfil this purpose, we have included and pre-configured a comprehensive set of parameters for possible modelling scenarios. At the current release, there are 6 separate profiles for the DM halo density and magnetic field strength and 5 separate profiles for gas density, with a straightforward method for adding further custom profile functions. There is also support for both WIMP annihilation and decay modes, with a default interface with the excellent PPPC4DMID~\cite{cirelliPPPCDMID2011} resource which provides particle spectra for all relevant SM products from these annihilations and decays. Outside of the typical input the user is expected to provide, the data structures used in the code are written with the intention of being simple to understand and comprise mostly of python dictionaries with explicit key-value pairs which makes the identification and manipulation of relevant variables as convenient as possible. 

When compared to existing open-source tools, \pdm{} provides several scientific and practical benefits, particularly when calculating radio emissions. In the \prx{} package, the calculation of the equilibrium electron distribution follows a semi-analytical prescription. This method results in an effective reduction of the magnetic field strength at the centre of the halo, while allowing diffusing electrons to travel further throughout the halo. This effect can thus underestimate the radio emissivity in these regions, particularly in compact objects with steep magnetic field profiles and small diffusion regions. In regards to practical use cases, the solution method adopted by \pdm{} also not only provides quicker computation times, but is more robust to artefacts associated with integration over the diffusion zone. We note that the \pkg{GALPROP} package is able to calculate the equilibrium electron distribution using a similar numerical solver to the one used here, thus also avoiding the issues mentioned above. However, it does not provide the host of auxiliary functions related to determining radio emissions from WIMPs that are included in \pdm{}, and has been designed primarily for use with galaxies that have a similar morphology to the MW, which limits its usefulness when considering a wide range of astrophysical targets and the observations associated with them.   

The aspects of the code mentioned above thus allow the indirect emissions from any DM halo target, from small dSph satellite galaxies to large galaxy clusters, to be used in the ongoing effort by the astronomy sector to complement terrestrial DM experiments. This is a vital aspect to indirect detection experiments, as the current generation of astronomical observatories are already producing huge, high-resolution and high-sensitivity datasets from a large number of targets, through both directed observations and survey programmes. The use of this data in DM searches thus requires not only accurate modelling of the emission process, but also the quick and practical computation provided by \pdm{} to keep up with the available data. The \pdm{} package has already been utilised successfully in a recent indirect detection study~\cite{lavisRadiofrequencyWIMPSearch2023} using MeerKAT Galaxy Cluster Legacy Survey~\cite{knowlesMeerKATGalaxyCluster2022} observations of several radio-faint galaxy clusters, setting stringent constraint on the WIMP annihilation cross-section. The open-source nature of this code should also allow for extra features to be added as they are developed by the community, so that this tool may come to play an increasingly helpful role in the ongoing hunt for DM in our universe.

\section{Acknowledgments}
  We thank Andrei Egorov for helpful discussions and suggestions regarding some calculations. GB acknowledges funding from the National Research Foundation of South Africa under the Thuthuka grant no. 117969. The work of MS was supported by the National Research Foundation
  of South Africa (Bursary No. 112332). This work makes use of the following software packages and code libraries: \texttt{AstroPy}~\cite{astropycollaborationAstropyProjectSustaining2022}, \texttt{SciPy}~\cite{scipy2020}, \texttt{NumPy}~\cite{numpy2020}, \texttt{SymPy}~\cite{meurer_sympy_2019} and \texttt{Matplotlib}~\cite{huntermatplotlib2007}.

\section*{Data availability statement}
Supplementary data used in this work is hosted in the Zenodo repository located at the doi:\url{https://doi.org/10.5281/zenodo.13312389}. The source code of the software package presented in this work is available through the GitHub repository \url{https://github.com/Hyperthetical/DarkMatters/}.

\appendix
\section{Solving the diffusion-loss equation}\label{sec:solution_methods}

The standard, spherically symmetric diffusion-loss equation used to describe the evolution of cosmic rays in a magnetic field is given by 
\begin{align}\label{eqn:diffusion_loss_app}
	\pdv{\psi(r,E)}{t} = \nabla\cdot(D(r,E)\nabla\psi(r,E)) + \pdv{}{E}\left(b(r,E)\psi(r,E)\right) + Q(r,E) \, .
\end{align}
Several techniques to solve this second-order partial differential equation for the equilibrium distribution of electrons exist in the literature. These include the semi-analytical methods of~\cite{colafrancescoMultifrequencyAnalysisNeutralino2006} (used in a multitude of WIMP indirect detection studies and the method of choice in the \pkg{RX-DMFIT} software package~\cite{mcdanielMultiwavelengthAnalysisDark2017}), and the method developed in~\cite{vollmannUniversalProfilesRadio2021} (also recently used in indirect WIMP searches~\cite{vollmannRadioConstraintsDark2020,gajovicWeaklyInteractingMassive2023}). The \pkg{DarkMatters} package employs a set of numerical techniques to solve this equation. Some of these techniques were inspired by existing codebases, like the \pkg{GALPROP}~\cite{strongPropagationCosmicrayNucleons1998} and \pkg{DRAGON(2)}~\cite{evoliCosmicrayPropagationDRAGON22017} packages, while others have been developed in this work. We also note that the authors of~\cite{regisLocalGroupDSph2015a} have used a similar numerical technique in a series of WIMP indirect detection studies~\cite{regisLocalGroupDSph2014,regisDarkMatterReticulum2017,regisEMUViewLarge2021}. 

The numerical solution used here is based on the Crank-Nicolson (CN) finite-differencing of Equation~\eqref{eqn:diffusion_loss_app}. The discretised operators of each dimension, namely the spatial and energy (momentum) dimensions, are then solved individually and alternatively in a technique known as Operator Splitting (OS), which leads to a drastic reduction in the computational requirements of the solution. Further, the full spatial dependence of magnetic field and gas density profiles that are relevant to the evolution of the system can in this case be considered at each step of the modelling process, unlike in semi-analytical methods. 

Firstly, we take Equation~\eqref{eqn:diffusion_loss_app} and make the variable transformations $\tr = \log_{10}(r/r_0)$ and $\tE = \log_{10}(E/E_0)$, where $r_0$ and $E_0$ are characteristic scaling values. In \pkg{DarkMatters}, the defaults for these values are $r_s$, the scale radius of the DM halo, and $1\,\mathrm{GeV}$, respectively. With the aforementioned assumption of spherical symmetry, the diffusion-loss equation then becomes
\begin{equation}\label{eqn:diffusion_loss_transformed}
    \pdv{\psi}{t} = \dfrac{10^{-2\tr}}{(\ln(10)r_0)^2}\left(\left[\ln(10)2D + \pdv{D}{\tr}\right]\pdv{\psi}{\tr} + D \pdv[2]{\psi}{\tr}\right) + \dfrac{10^{-\tilde{E}}}{\ln(10)E_0}\left( \pdv{b}{E}\psi + b\pdv{\psi}{E}\right) + Q\,,
\end{equation}
where $\psi = \psi(\tr,\tE)$. This transformation of variables results in function evaluations and discrete grid points that are more convenient to handle, given that the relevant physical scales can encompass many orders of magnitude. To discretise Equation~\eqref{eqn:diffusion_loss_transformed}, we make use of the CN scheme, which is described in~\cite{pressNumericalRecipesArt2007}. This scheme uses the average of explicit and implicit differencing terms, which has the benefit of having the unconditional stability of implicit methods as well as the second-order accuracy for small-scale effects. A representation of this scheme applied to a one-dimensional diffusion-loss equation can be written as in~\cite{strongPropagationCosmicrayNucleons1998}, 
\begin{equation}\label{eqn:cn_alpha}
	\dfrac{\psi_i^{n+1}-\psi_i^n}{\Delta t} = \dfrac{\alpha_1 \psi_{i-1}^{n+1} - \alpha_2 \psi_{i}^{n+1} + \alpha_3 \psi_{i+1}^{n+1}}{\strut 2\Delta t} + \dfrac{\alpha_1 \psi_{i-1}^{n} - \alpha_2 \psi_{i}^{n} + \alpha_3 \psi_{i+1}^{n}}{\strut 2\Delta t} + Q_i\,.  
\end{equation}
Here the functions have been discretised over the grid with spatial and temporal indices given by $i$ and $n$, and the grid spacing is $\Delta t = t^{n+1}-t^n$. The $\alpha$ coefficients are generalised coefficients containing some combination of the diffusion and energy loss functions and grid spacings which will be defined below. The form of Equation~\eqref{eqn:cn_alpha} shows the use of both explicit ($n+1$ terms) and implicit ($n$ terms) differencing techniques. This equation is also commonly written in the form 
\begin{align}\label{eqn:cn_isolated}
	-\dfrac{\alpha_1}{2}\psi^{n+1}_{i-1} + \left(1+\dfrac{\alpha_2}{2}\right)\psi^{n+1}_{i} - \dfrac{\alpha_3}{2}\psi^{n+1}_{i+1} = Q_i\Delta t + \dfrac{\alpha_1}{2}\psi^{n}_{i-1} + \left(1- \dfrac{\alpha_2}{2}\right)\psi^{n}_{i} + \dfrac{\alpha_3}{2}\psi^{n}_{i+1}\,,
\end{align}
where implicit and explicit terms have been grouped to yield the overall updating equation
\begin{equation}\label{eqn:updating}
	A\psi^{n+1} = B\psi^n + Q \,.
\end{equation}
The matrices in this updating equation have tri-diagonal forms which contain the $\alpha$-coefficients. The tri-diagonal nature of these matrices is an important factor in the computation of solutions, as specialised algorithms have been developed that perform the calculation more efficiently than fully populated matrices. However, when generalising this scheme to multiple dimensions, the tri-diagonality of the matrices is lost. The OS method is thus used to retain the benefit of the tri-diagonal matrix structure. Representing the diffusion-loss equation as
\begin{equation}
	\pdv{\psi}{t} = \mathcal{L}\psi\,,
\end{equation}
where $\mathcal{L}\psi$ is a linear combination of independent operators, then each linearly independent operator can be applied in turn during an iterative solution algorithm, as long as each operator has a valid finite differencing scheme in the absence of all others. In this case, the operators for the spatial and energy dimensions in Equation~\eqref{eqn:diffusion_loss_transformed} can be split and written as
\begin{equation}
	\mathcal{L}_r = \dfrac{10^{-2\tr}}{(10\ln(10)r_0)^2}\left(2D\dfrac{\partial}{\partial \tr} + \dfrac{\partial D}{\partial \tr}\dfrac{\partial}{\partial \tr} + D \dfrac{\partial^2}{\partial \tr^2}\right) \,
\end{equation}
and
\begin{equation}
	\mathcal{L}_E = \pdv{b}{E} + b\pdv{E} \,.
\end{equation}
Representing the finite differencing schemes of these operators symbolically as $\Psi_r$ and $\Psi_E$, we obtain the following 
\begin{equation}
	\mathcal{L}_r \rightarrow \Psi_r = \dfrac{10^{-2\tr_i}}{(10\ln(10)r_0)^2}\left[\dfrac{\psi_{i+1}-\psi_{i-1}}{2\Delta \tr}\left(\ln(10)2D + \pdv{D}{\tr}\right) + \dfrac{\psi_{i+1}-2\psi_{i}+ \psi_{i-1}}{\Delta \tr^2}D\right] \,
\end{equation} 
and
\begin{equation}
	\mathcal{L}_E \rightarrow \Psi_E = \dfrac{10^{-\tE_j}}{(E_0\ln(10))}\left[\dfrac{b\psi_{j+1}-b\psi_{j}}{\Delta \tE}\right] \,.
\end{equation}
We have only considered the upstream case for the energy dimension, as in \pkg{GALPROP} and~\cite{regisLocalGroupDSph2015a}, under the assumption that electrons only lose energy. In the above schemes the $\Delta$ quantities have their usual meanings and the subscripts $i$ and $j$ have been reserved for the spatial and energy grids, respectively. The overall updating scheme for the points $n$ to $n+1$, using the OS method, can then be expressed symbolically as
\begin{align}\label{eqn:os}
	\psi^{n+1/2} &= \Psi_r(\psi^n,\Delta t/2) \nonumber \\
	\psi^{n+1} &= \Psi_E(\psi^{n+1/2},\Delta t/2).
\end{align}
The computation of Equation~\eqref{eqn:os} requires the solution of large matrix equations at each iteration, which can draw significant compute power, especially in the case of high-resolution grids. To mitigate these computational requirements, we describe below a technique to reduce the number of iterations and method calls necessary for the algorithm to run, significantly speeding up the runtime of the code. Now, the discretised 2-dimensional electron distribution $\psi_{i,j}$ is represented by a 2-dimensional array in the code of form $\psi \in \mathbb{R}^{I\times J}$, where the 2 spatial indices $i,j$ run from $0$ to $I,J$, respectively. This function is flattened to a single column vector in the operation  $\psi \in \mathbb{R}^{I\times J} \rightarrow \psi \in \mathbb{R}^{IJ \times 1}$. There are two ways to perform this operation, and in each dimension ($\tr$ or $\tE$) the flattened vector will have blocks of sub-matrices that are composed solely of function values that are independent of the alternate dimension. For example, when performing the $\tr$ step in the algorithm, the first $I$ elements in the column vector will compose one block of values that are independent of $\tE$. Consequently, each of the matrices $A$ and $B$ in the general updating system (Equation~\eqref{eqn:updating}) will then also be composed of an equal number of sub-matrix blocks that act on the independent blocks in $\psi$. We represent these tridiagonal block matrices as $\mathcal{M}$, and they have the following form (derived from Equation~\ref{eqn:cn_isolated}):
\begin{equation}
	\mathcal{M}({\tr},E_j) = 
	\begin{pmatrix}
		\left(1+\frac{\alpha_2(\tr_0,\tE_j)}{2}\right) & - \frac{\alpha_3(\tr_0,\tE_j)}{2} &  &  & \\
		-\frac{\alpha_1(\tr_1,\tE_j)}{2} & \left(1+\frac{\alpha_2(\tr_1,\tE_j)}{2}\right)  &  &  & \\
		& & \ddots & & \\
		& & & \left(1+\frac{\alpha_2(\tr_{I-1},\tE_j)}{2}\right) & - \frac{\alpha_3(\tr_{I-1},\tE_j)}{2}\\
		& & & -\frac{\alpha_1(\tr_{I},\tE_j)}{2} & \left(1+\frac{\alpha_2(\tr_{I},\tE_j)}{2}\right)\\
	\end{pmatrix}
\end{equation}
and similarly for $\mathcal{M}(\tr_i,\tE)$. With this definition, the full $A$ matrices in each step of the algorithm then resemble:
\begin{align}\label{eqn:blockr}
	A_{\tr} &= 
	\begin{pmatrix}
		\mathcal{M}(\tr,\tE_0) & & & & & 0\\
		& \mathcal{M}(\tr,\tE_1) & & & &\\
		& & \ddots & & &\\
		& & &\mathcal{M}(\tr,\tE_{J-1})& \\
		0 & & & & & \mathcal{M}(\tr,\tE_{J})\\
	\end{pmatrix}
\end{align}
for the spatial dimension and
\begin{align}\label{eqn:blockE}
	A_{\tE} &= 
	\begin{pmatrix}
		\mathcal{M}(\tr_0,\tE) & & & & & 0\\
		& \mathcal{M}(\tr_1,\tE) & & & &\\
		& & \ddots & & &\\
		& & &\mathcal{M}(\tr_{I-1},\tE)& \\
		0 & & & & & \mathcal{M}(\tr_{I},\tE)\\
	\end{pmatrix}
\end{align}
for the energy dimension (the $B$ matrices are constructed similarly). The total size of each matrix is then the total size of each block matrix times the number of block matrices, for example $\mathcal{M}(\tr,E_j) \in \mathbb{R}^{I\times I}$ and $A_{\tr} \in \mathbb{R}^{IJ\times IJ}$. Each half-step of Equation~\eqref{eqn:updating} can now be solved with a single matrix equation solution function call, instead of $I$ or $J$ times.

The forms of the matrices in Equation~\eqref{eqn:blockr} and~\eqref{eqn:blockE} are clearly extremely sparse. In most practical scenarios, the sparsity index (calculated by $\mathcal{S} = (N - N_{nz})/N\times100\%\,,$ where $N$ and $N_{nz}$ are the total number and number of non-zero elements in the matrix) is~$> 99.9\%$. Thus, when working with these matrices in the code we have made use of the \texttt{scipy.sparse} Python module, which stores the above matrices in various memory-efficient storage configurations. In particular, the matrices are stored in the `csr' (compressed-sparse-row) storage format, which allows for fast array vector operations. Since the solution of the matrix equation is required at every iteration of the algorithm, the efficiency of the overall numerical solution is directly linked to the efficiency of this library's sparse matrix implementation. 

The forms of the discretised operators that appear in the OS scheme in Equation~\eqref{eqn:os} are now accessible through the method of equating coefficients, and we find:
\begin{equation}\label{eqn:alpha_r}
	\Psi_{\tr}:
	\begin{cases}
		\dfrac{\alpha_1}{\Delta t} &= C_{\tr}\left(-\dfrac{\ln(10)D(\tr_i,E_j)+\eval{\pdv{D}{\tr}}_{\tr_i,E_j}}{2\Delta\tr} + \dfrac{D(\tr_i,E_j)}{\Delta\tr^2}\right)\\[1.5em]
		\dfrac{\alpha_2}{\Delta t} &= C_{\tr}\left(\dfrac{2D(\tr_i,E_j)}{\Delta\tr^2}\right)\\[1em]
		\dfrac{\alpha_3}{\Delta t} &= C_{\tr}\left(\dfrac{\ln(10)D(\tr_i,E_j)+\eval{\pdv{D}{\tr}}_{\tr_i,E_j}}{2\Delta\tr} + \dfrac{D(\tr_i,E_j)}{\Delta\tr^2}\right)
	\end{cases}
\end{equation}
for the spatial dimension and 
\begin{equation}\label{eqn:alpha_E}
	\Psi_{E}:
	\begin{cases}
		\dfrac{\alpha_1}{\Delta t} &= 0\\[1em]
		\dfrac{\alpha_2}{\Delta t} &= C_{\tE}\dfrac{b_j}{\Delta E}\\[1em]
		\dfrac{\alpha_3}{\Delta t} &= C_{\tE}\dfrac{b_{j+1}}{\Delta E}
	\end{cases}
\end{equation}
for the energy, where the coefficients are defined by $C_{\tr}=\dfrac{10^{-2\tr_i}}{(\ln(10)r_0)^2}$ and $C_{\tE} = \dfrac{10^{-\tilde{E}_j}}{\ln(10)E_0}$. We use the DM electron source function $Q(\tr,\tE)$ as the initial condition, and we apply the following Dirichlet and Neumann boundary conditions: 
\begin{align}
	\psi(\tr = \tr_{\mathrm{max}}) &= 0 \,,\nonumber \\ 
	\pdv{\psi}{\tr}\, (\tr = \tr_{\mathrm{min}}) &= 0 \,.
\end{align}
The Dirichlet condition corresponds to the free-streaming of the electrons out of the halo at the maximum diffusion radius, and is enforced simply by setting all values of $\psi$ at $\tr=\tr_{\mathrm{max}}$ to 0 at each iteration of the OS method. The Neumann condition is satisfied by the calculation of new $\alpha$-coefficients for the points $\tr=\tr_{\mathrm{min}}$, which represents the smallest sampled radius in the spatial grid. Analogously to Equation~\eqref{eqn:alpha_r}, we find
\begin{equation}
	\Psi_{\tr=\tr_{\mathrm{min}}}:
	\begin{cases}
		\dfrac{\alpha_1}{\Delta t} &= 0 \\[1em]
		\dfrac{\alpha_2}{\Delta t} &= C_{\tr}\left(\dfrac{4D(\tr_i,\tE_j)}{\Delta\tr^2}\right)\\[1em]
		\dfrac{\alpha_3}{\Delta t} &= C_{\tr}\left(\dfrac{4D(\tr_i,\tE_j)}{\Delta\tr^2}\right)
	\end{cases}
\end{equation}
As described in Section~\ref{sec:implementation}, the values of $\tr_{\mathrm{max}}$ and $\tr_{\mathrm{min}}$ are configurable in the input parameter file. 

Finally, we discuss the conditions which determine the continuation and eventual termination of the algorithm to find the equilibrium electron distribution. These conditions are closely related to the various physical timescales that are involved in the problem, which we define as follows. The timescales of energy-loss and diffusion are $\tau_{E} = \tE/b(\tE)$ and $\tau_{D} = \Delta \tr^2/D(\tr,\tE)$ respectively, and the timescale of changes to the $\psi$ distribution are $\tau_{\psi} = \psi/{\left\vert\left(\pdv{\psi}{t}\right)\right\vert}$ which we estimate using a simple first-order Forward-Time difference. We have implemented two separate methods to advance each step of the updating algorithm given in Equation~\eqref{eqn:os}, each inspired by the techniques developed in the \pkg{GALPROP} package (see~\cite{strongPropagationCosmicrayNucleons1998} or the Explanatory Supplement\footnote{\url{https://galprop.stanford.edu/code.php?option=manual}}). 

In the first method, called the `accelerated' (ACC) method, the initial timestep $\Delta t_i$ is chosen to be larger than the timescales of both energy-loss and diffusion. Then, after a set number of iterations, $\Delta t$ is reduced by a factor of 2, after which the entire process repeats until $\Delta t$ reaches a minimum value. This timestep-changing technique allows for all relevant physical scales to be included in the final solution, while reducing the total number of iterations required to find an accurate solution. The motivation for this method arises from considering a variety of astrophysical structures and scenarios, which could have a wide range of timescales related to energy-loss or diffusion impacting the evolution of the electron distribution. If the timestep $\Delta t$ is small enough to capture the short timescale effects, converging on the final equilibrium distribution could require a huge number of iterations. Conversely, while needing only relatively few iterations, large timesteps can lead to inaccurate final solutions as the impact from short-timescale processes are lost during each iteration. If the variables governing the ACC method (through the parameters \verb|os_delta_ti|, \verb|os_max_steps| and \verb|os_delta_t_min|) are chosen appropriately, the dynamic nature of the timestep value solves both of the above issues during a typical run, and it is thus the recommended method when using \pdm{}.

The other possible method has no cap on the number of iterations, and has a constant step size (CSS) for the timestep $\Delta t$. In this method a tolerance $T$ can be set for the relative difference between the electron distribution at subsequent iterations, and if the relative difference is below this tolerance, i.e. $\left\vert 1-\psi^{n-1}/\psi^n \right\vert < T$, then the algorithm is considered to be stable. This condition ensures that the changes in the distribution over all the grid points (which are initially large) approach zero instead of diverging or oscillating due to potential numerical effects. Because of the range of timescales discussed above, this method typically takes much longer to find the solution as the ACC method. However, we have also utilised this method to perform a set of benchmarking tests that have strict criteria for the final convergence of the algorithm to the equilibrium distribution. In these tests, we use a very small timestep ($\Delta t < \min\{\tau_E,\tau_D\}$) with no cap on the iteration number. Then, similarly to the tests performed in \pkg{GALPROP}, we signal convergence with the following condition: 
\begin{equation}
	\pdv{\psi}{t} = 0 \nonumber
\end{equation}
\begin{equation}\label{eqn:benchmark}
	\Rightarrow \dfrac{\psi^{n-1} - \psi^{n}}{\Delta t} = 0 \,,
\end{equation}
where Equation~\eqref{eqn:benchmark} must hold for all points in the electron distribution. While ensuring the maximum level of accuracy in the final solution, these runs are most useful as a comparative tool, as the number of iterations required to satisfy the above condition is usually prohibitive for practical purposes. As the relative difference in the electron distributions between iterations also tends to rapidly diminish at large iteration numbers, these final steps can in general be avoided without significant risk of a loss of accuracy. 

Outside of benchmarking and accuracy tests, the general condition for convergence to the equilibrium distribution is set by
\begin{equation}
    \tau_{\psi} > \max\{\tau_E,\tau_D\} \,,
\end{equation}
which ensures that the timescale of changes to the electron distribution is larger than the timescale of any physical effects involved. 

\bibliographystyle{elsarticle-num}
\bibliography{references}

\end{document}